\documentclass[a4paper,preprintnumbers,floatfix,twocolumn,aps,prb,unsortedaddress,superscriptaddress]{revtex4-2}

\usepackage[para]{threeparttable} 
\usepackage{amsmath}
\usepackage{graphicx} 
\usepackage{booktabs}
\usepackage{subcaption}
\usepackage{color}
\usepackage[normalem]{ulem}  
\usepackage{url}
\usepackage{hyperref}
\usepackage{miller}
\usepackage{gensymb}

\usepackage[utf8]{inputenc}

\begin{document}

\title{Machine-learning interatomic potential for radiation damage and defects in tungsten}

\author{J. Byggmästar}
\thanks{Corresponding author}
\email{jesper.byggmastar@helsinki.fi}
\affiliation{Department of Physics, P.O. Box 43, FI-00014 University of Helsinki, Finland}
\author{A. Hamedani}
\affiliation{Department of Physics, P.O. Box 43, FI-00014 University of Helsinki, Finland}
\affiliation{Engineering Department, Shahid Beheshti University, G.C, P.O. Box 1983969411, Tehran, Iran}
\author{K. Nordlund}
\affiliation{Department of Physics, P.O. Box 43, FI-00014 University of Helsinki, Finland}
\author{F. Djurabekova}
\affiliation{Department of Physics, P.O. Box 43, FI-00014 University of Helsinki, Finland}
\affiliation{Helsinki Institute of Physics, Helsinki, Finland}

\date{\today}

\begin{abstract}
{We introduce a machine-learning interatomic potential for tungsten using the Gaussian Approximation Potential framework. We specifically focus on properties relevant for simulations of radiation-induced collision cascades and the damage they produce, including a realistic repulsive potential for the short-range many-body cascade dynamics and a good description of the liquid phase. Furthermore, the potential accurately reproduces surface properties and the energetics of vacancy and self-interstitial clusters, which have been long-standing deficiencies of existing potentials. The potential enables molecular dynamics simulations of radiation damage in tungsten with unprecedented accuracy.}
\end{abstract}

\maketitle

\section{Introduction}
\label{sec:intro}

Tungsten is considered to be the only viable material for the highest heat load parts of an energy-producing fusion reactor. High-energy neutrons emitted from the fusion plasma initiate collision cascades in the wall material, leading to permanent damage. Understanding the radiation-induced microstructural changes and evolution is therefore a topic of active research~\cite{rieth_recent_2013,zinkle_designing_2014,nordlund_primary_2018}. Achieving a comprehensive understanding of the radiation damage requires a combined effort of experimental measurements and theoretical modelling. Atomistic simulations using classical molecular dynamics (MD) have been a fruitful tool for understanding the atom-level damage production that is beyond reach of experimental time and length scales, such as the formation and morphology of radiation-induced defects~\cite{sand_high-energy_2013,byggmastar_collision_2019}. The accuracy of MD relies, however, completely on the accuracy of the interatomic potential, which is typically a relatively simple analytical function fitted to reproduce a selected set of material properties.

Simulating collision cascades and the damage they produce is a particularly challenging task for the interatomic potential. The material experiences a number of atom-level changes during the evolution of a cascade, including many-body atom collisions, localised melting, rapid recrystallisation with extreme temperature and pressure gradients, and defect clustering. Describing all these aspects pushes the interatomic potential to (and often beyond) its limits, and different potentials can give widely different results~\cite{nordlund_defect_1998,sand_non-equilibrium_2016,byggmastar_effects_2018,stoller_impact_2016}. For radiation damage in tungsten and other metals, the embedded atom method potentials~\cite{daw_embedded-atom_1984}, and to a lesser extent Tersoff-like bond-order potentials~\cite{tersoff_new_1988}, have been particularly successful~\cite{ackland_improved_1987,derlet_multiscale_2007,marinica_interatomic_2013,mason_empirical_2017,chen_new_2018,juslin_analytical_2005,ahlgren_bond-order_2010}. Nevertheless, several key properties of tungsten have remained challenging to capture. For example, many potentials fail to reproduce the relative stability of dislocation loops~\cite{fikar_nano-sized_2018,byggmastar_collision_2019,chen_new_2018}, which leads to large differences especially in the damage produced by overlapping cascades~\cite{byggmastar_collision_2019,fellman_radiation_2019}. Most existing potentials suitable for radiation damage in tungsten also fail to reproduce the binding of small vacancy clusters, such as the peculiar repulsion of the di-vacancy~\cite{ventelon_ab_2012,heinola_stability_2018}. Furthermore, the majority of tungsten potentials consistently underestimate the surface energy by up to 30--40\%, and struggle to reproduce the order of stability of different surfaces~\cite{bonny_many-body_2014,mason_empirical_2017}. All of these deficiencies can be attributed to the limited flexibility of the fixed functional form of the potentials, whereby some properties often can be impossible to reproduce without sacrificing more important properties.

In the past decade, various forms of machine-learning interatomic potentials have become increasingly popular~\cite{behler_generalized_2007,bartok_gaussian_2010,thompson_spectral_2015,shapeev_moment_2016,glielmo_efficient_2018,zhang_deep_2018,behler_perspective_2016}. The main advantage of using machine learning to construct potentials is that a fixed analytical form is not assumed, which results in flexible potentials that can describe virtually any material and their properties. Additionally, machine-learning potentials can be systematically improved towards the accuracy of the training data by increasing the degrees of freedom of the model. The main practical limitation of machine-learning potentials is the computational speed, which is typically several orders of magnitude lower than analytical potentials~\cite{zuo_performance_2019}. However, more efficient implementations and optimisations will likely reduce the computational cost of machine-learning potentials significantly, as recently demonstrated in Ref.~\cite{caro_optimizing_2019}. In this work, we employ the Gaussian Approximation Potential (GAP) framework~\cite{bartok_gaussian_2010,bartok_representing_2013,bartok_gaussian_2015} to develop a potential for tungsten, with particular focus on radiation damage. The rest of the article is structured as follows. In section~\ref{sec:potential} we introduce the mathematical structure of the potential. Computational details are summarised in section~\ref{sec:comp}. In section~\ref{sec:training} we describe in detail the training strategy along with the contents of the training database. We subject the trained potential to extensive benchmarking for validation in section~\ref{sec:results}, followed by an outlook and concluding remarks in section~\ref{sec:concl}.

\section{Potential details}
\label{sec:potential}

The total energy of an atom $i$ in the GAP formalism is evaluated using Gaussian process regression~\cite{rasmussen_gaussian_2006,bartok_gaussian_2010}, and can be written as a sum over basis (kernel) functions
\begin{equation}
    E_i = \delta^2 \sum_s^M \alpha_s K (\textbf{q}_i, \textbf{q}_s),
\end{equation}
where $s$ loops over a set of $M$ representative atoms from the training database. $\delta$ sets the scale and range of energies to be trained. $K$ is the kernel function, which acts as a measure of similarity between the atomic environment of the known atom $s$ and the desired atom $i$. The local atomic environments are quantified by the descriptor $\textbf{q}$. The keys to achieving good accuracy lie in the choices of kernel functions and descriptors, as well as in a clever construction of the training database. The energies (along with forces and possibly virial stresses) from the training database are learned by optimising the coefficients $\alpha_s$ given by the solution of a regularised least-squares problem~\cite{bartok_machine_2018}. Regularisation is applied by supplying weights in proportion to the expected errors of the training data, $\sigma_\nu$ (which should include both the uncertainties of the training data and errors due to assuming a finite range of the GAP). For a more detailed description of the mathematical framework of GAP, we refer to Refs.~\cite{bartok_gaussian_2015,bartok_machine_2018}.

When training the GAP, we use a combination of two descriptors with associated kernels. A simple two-body descriptor (i.e. the interatomic distance) with the Gaussian-like squared exponential kernel is used to capture the majority of the interatomic bond energies. As is typical for GAP models, the many-body interactions are described by the Smooth Overlap of Atomic Positions (SOAP) kernel~\cite{bartok_representing_2013}. We tried including a three-body descriptor, but found that the accuracy was only marginally increased and therefore rely on SOAP for capturing all many-body effects. The mathematical background of SOAP has been extensively described in Ref.~\cite{bartok_representing_2013} and will not be repeated in detail here. Shortly, SOAP compares two atomic environments by integrating the overlap of their smeared atomic densities, as obtained by placing Gaussian functions centred on each atom position within the cutoff radius. In addition to the GAP, we use an external pair potential to take care of the extreme repulsion at short interatomic distances, as discussed in detail below. The total energy of a system of $N$ atoms then reads
\begin{align}
 \begin{split}
    E_\mathrm{tot} &= \sum_{i<j}^N V_\mathrm{pair}(r_{ij}) + \sum_i^N E_\mathrm{GAP} \\
    &= \sum_{i<j}^N V_\mathrm{pair}(r_{ij}) \\
    &+ \delta_\mathrm{2b}^2 \sum_i^N \sum_s^{M_\mathrm{2b}} \alpha_{s, \mathrm{2b}} K_\mathrm{2b} (\textbf{q}_{i, \mathrm{2b}}, \textbf{q}_{s, \mathrm{2b}}) \\
    &+ \delta_\mathrm{mb}^2 \sum_i^N \sum_s^{M_\mathrm{mb}} \alpha_{s, \mathrm{mb}} K_\mathrm{mb} (\textbf{q}_{i, \mathrm{mb}}, \textbf{q}_{s, \mathrm{mb}}),
  \end{split}
\end{align}
where the 2b and mb subscripts are used to separate the two-body and many-body (SOAP) terms. The hyperparameters associated with the two descriptor terms used when training the GAP are listed in Tab.~\ref{tab:hyper}, along with short descriptions for each parameter. The interaction range of both descriptor terms is limited by a 5 Å cutoff radius. We tested several cutoff values in the 3--7 Å range, and found 5 Å to be a reasonable choice. The values for $N_\mathrm{sparse}$, $n_\mathrm{max}$, $l_\mathrm{max}$ and $\delta$ were chosen following systematic convergence tests. Nevertheless, we note that the accuracy of the GAP is not particularly sensitive to the exact hyperparameter values listed in Tab.~\ref{tab:hyper}.

The internuclear repulsion at extremely short distances is accounted for by the external pair potential in the form of a screened Coulomb potential
\begin{equation}
    V_\mathrm{pair} (r_{ij}) = \frac{1}{4\pi \varepsilon_0} \frac{Z_i Z_j e^2}{r_{ij}} \phi (r_{ij}/a) f_\mathrm{cut} (r_{ij}),
\end{equation}
where
\begin{equation}
    a = \frac{0.46848}{Z_i^{0.23} + Z_j^{0.23}}.
\end{equation}
The function is identical to the universal Ziegler-Biersack-Littmarck (ZBL) potential~\cite{ziegler_stopping_1985}, but the screening function $\phi (x)$ is refitted specifically for W-W repulsion using the all-electron DFT-DMol data from Ref.~\cite{nordlund_repulsive_1997}. The DFT-DMol calculations were optimised for the high-energy repulsive part and recently found to show excellent agreement with experiments~\cite{zinoviev_comparison_2017}. We refit the screening function for two reasons. First, we found that the ZBL potential for W-W is noticeably different than both the all-electron DFT-DMol data and our \textsc{vasp} data. Second, it is useful to have some freedom for tuning the pair potential, in order to ensure a smooth connection with the near-equilibrium energies and forces to be learned by the GAP. We accomplish this by making sure that the screened Coulomb potential smoothly joins and closely matches the repulsive energy and forces corresponding to the closest interatomic distances present in the training structures (see Fig.~\ref{fig:dimer}). Only the differences in energies and forces between the external pair potential and the training data need to be reproduced by the GAP. Hence, the GAP is taught to predict energies and forces close to zero for short interatomic distances, so that the screened Coulomb potential fully dictates the short-range dynamics, as desired. The fitted screening function is
\begin{align}
 \begin{split}
    \phi (x) &= 0.32825 \exp(-2.54931x) \\
    &+ 0.09219 \exp(-0.29182x) \\
    &+  0.58110 \exp(-0.59231x).
 \end{split}
\end{align}
The screened Coulomb potential is forced to zero by the cutoff function
\begin{equation}
    f_\mathrm{cut} (r) = 
 \begin{cases}
  1, & r \leq r_1 \\
  1 - \chi^3(6\chi^2 - 15\chi + 10), & r_1 < r < r_2 \\
  0, & r \geq r_2,\\
 \end{cases}
\label{eq:cutoff}
\end{equation}
where $\chi = (r - r_1) / (r_2 - r_1)$. The cutoff range is chosen as $r_1 = 1$ Å and $r_2 = 2.2$ Å, leaving the near-equilibrium bond energies to be fully machine-learned. The cutoff function is the same as in~\cite{perriot_screened_2013}, and is continuous at the end-points up to the second derivative. In practice, $V_\mathrm{pair}$ is tabulated and provided as input when training the GAP.

\begin{table}
 \centering
  \caption{Input parameters used when training the GAP. $R_\mathrm{cut}$: cutoff radius, $R_{\Delta \mathrm{cut}}$: width of cutoff region, $\delta$: energy scale, $N_\mathrm{sparse}$: number of sparse points (representative environments picked from the training structures), $n_\mathrm{max}$ and $l_\mathrm{max}$: limits of spherical harmonics used in SOAP,  $\sigma_\mathrm{atom}$: width of atomic Gaussians in SOAP, $\zeta$: exponent of SOAP kernel. For more details, see Ref.~\cite{bartok_gaussian_2015}.}
 \label{tab:hyper}
 \begin{tabular}{lll}
  \toprule
  & SOAP & Two-body  \\
  \midrule
  $R_\mathrm{cut}$ & 5 Å & 5 Å\\
  $R_{\Delta \mathrm{cut}}$ & 1 Å & 1 Å \\
  $\delta$ & 2 eV & 10 eV \\
  $N_\mathrm{sparse}$ & 4000 & 20 \\
  Sparse method & CUR & Uniform \\
  $n_\mathrm{max}$ & 8 & - \\
  $l_\mathrm{max}$ & 8 & - \\
  $\sigma_\mathrm{atom}$ & 0.5 Å & - \\
  $\zeta$ & 4 & - \\
  \bottomrule
 \end{tabular}
\end{table}

\section{Computational details}
\label{sec:comp}

The DFT training structures were calculated using \textsc{vasp}~\cite{kresse_ab_1993,kresse_ab_1994,kresse_efficiency_1996,kresse_efficient_1996} and the PBE GGA exchange-correlation functional~\cite{perdew_generalized_1996}. The 14 $5s^25p^65d^46s^2$ electrons were treated as valence electrons with the core electrons accounted for by the Projector-Augmented Wave (PAW) method~\cite{blochl_projector_1994,kresse_ultrasoft_1999} (the \texttt{W\_sv} PAW potential in \textsc{vasp} 5.4.4). The plane-wave cutoff energy was 500 eV and the Brillouin zone was integrated using Monkhorst-Pack grids~\cite{monkhorst_special_1976} with a consistent spacing between $k$-points for all cell sizes (using \texttt{KSPACING=0.15} Å$^{-1}$ in \textsc{vasp}, resulting in e.g. a $5\times 5\times 5$ grid for a 54-atom conventional bcc cell). A smearing of 0.1 eV by the first-order Methfessel-Paxton method~\cite{methfessel_high-precision_1989} was applied to help the convergence. The same settings were used for both the training and validation data.

The GAP was trained using \textsc{quip}~\cite{quip}. All molecular dynamics simulations were performed using \textsc{lammps}~\cite{plimpton_fast_1995} compiled with \textsc{quip} for GAP support~\cite{quip}. Phonon dispersion, nudged elastic band (NEB), and molecular statics calculations were performed within the Atomic Simulation Environment (ASE) framework~\cite{larsen_atomic_2017}. For calculations within the quasi-harmonic approximation, we used the \textsc{phonopy} code~\cite{togo_first_2015}.

\section{Training}
\label{sec:training}

\begin{table}
 \centering
  \caption{Structures included in the training database. $N_\mathrm{s}$ is the number of each structure type, $N_\mathrm{atoms}$ is the number of atoms in each structure, $N_\mathrm{atoms}^\mathrm{tot}$ is the total number of atoms of a given structure type, and $N_\mathrm{rep.}$ the number of representative atoms picked for the SOAP descriptor.}
 \label{tab:training_data}
  \begin{tabular}{lllllll}
   \toprule
    Structure type & $N_\mathrm{s}$ & $N_\mathrm{atoms}$ & $N_\mathrm{atoms}^\mathrm{tot}$ & $N_\mathrm{rep.}$ \\
   \midrule
   Isolated atom & 1 & 1 & 1 & 1 \\
   Dimer & 13 & 2 & 26 & 13 \\
   Distorted bcc unit cells & 2496 & 1--2 & 2996 & 69 \\
   Distorted other crystals: &   \\
   \quad fcc & 100 & 1 & 100 & 35 \\
   \quad hcp & 100 & 2 & 200 & 24 \\
   \quad A15 & 100 & 8 & 800 & 141 \\
   \quad C15 & 100 & 6 & 600 & 142 \\
   \quad dia & 100 & 2 & 200 & 66 \\
   \quad sc & 100 & 1 & 100 & 59 \\
   High-$T$ bcc & 20 & 54 & 1080 & 23 \\
   Vacancies: &  \\
   \quad single vacancy & 200 & 53 & 10600 & 201 \\
   \quad di-vacancies & 10 & 118 & 1180 & 25 \\
   \quad tri-vacancies & 15 & 117 & 1755 & 46 \\
   Self-interstitials (SIAs): & \\
   \quad single SIAs & 32 & 121 & 3872 & 113 \\
   \quad di-SIAs & 15 & 122, 252 & 2350 & 93 \\
   bcc surfaces  &  \\
   \quad \hkl(100) & 45 & 12 & 540 & 27 \\
   \quad \hkl(110) & 45 & 12 & 540 & 9 \\
   \quad \hkl(111) & 43 & 12 & 516 & 50 \\
   \quad \hkl(112) & 45 & 12 & 540 & 34 \\
   Liquids & 46 & 128 & 5760 & 1937 \\   
   Disordered surfaces & 24 & 128, 144 & 3264 & 461 \\
   Short-range & 100 & 53--55 & 5390 & 431 \\
   \midrule
   All & 3749 & & 42410 & 4000 \\
   \bottomrule
  \end{tabular}
\end{table}

Fitting an interatomic potential suitable for all aspects of radiation damage is a challenging task. The potential must be able to reproduce a wealth of properties and atomic geometries that might be encountered during the evolution of a collision cascade and the subsequent recrystallisation of the molten cascade core. Among the most important properties are the energy landscape and relative stability of various defects, from single vacancies and self-interstitial atoms (SIAs) to defect clusters. The potential must also reproduce realistic short-range dynamics defined by the repulsive part of the potential. Additionally, melting and recrystallisation as well as the structure of the liquid phase at various densities should be well described, in order to reproduce a realistic atomic mixing during the heat spike of a collision cascade. Furthermore, if surface irradiation is of interest, the energetics of perfect and damaged surfaces must be considered. No single existing potential is able to capture all of these aspects, and it is our goal to construct a training database of structures that captures all of the above-mentioned properties. Previously, Szlachta et al. trained a GAP for tungsten~\cite{szlachta_accuracy_2014} that excellently reproduces the properties of screw dislocations and vacancies. It was not, however, trained to self-interstitial atoms or the liquid phase and did not include a realistic repulsive potential, and is therefore not applicable to radiation damage simulations.

Table~\ref{tab:training_data} lists the types of structures included in the training database. The isolated atom is included to reproduce the correct cohesion. The elastic response of bcc tungsten was trained using randomly distorted unit cells. Part of these structures were taken from the training data of the previous W GAP~\cite{szlachta_accuracy_2014}. As we are interested in physics far from equilibrium, we included unit cells with large elastic distortions (with volumes up to about $\pm30$\% of the equilibrium volume). Similar elastically and randomly distorted unit cells were prepared for the fcc, hcp, A15, C15, diamond cubic, and simple cubic crystal structures. These serve to expose the GAP to additional high-symmetry atomic environments. 

Finite-temperature lattice vibrations were accounted for by including snapshots from MD simulations at 1000 K with three different volumes. The MD simulations were performed using an early version of our GAP, trained only to a small initial part of the training database. The structures containing a single vacancy were taken from~\cite{szlachta_accuracy_2014} (although only half of them were used in training while the other half were left for validation). We also added various di-vacancy and tri-vacancy structures to provide better transferability to clusters of multiple vacancies. Furthermore, we prepared SIAs in the common high-symmetry configurations in bcc: \hkl<111>, \hkl<110>, and \hkl<100> dumbbells, and atoms in the octahedral and tetrahedral sites. 

We checked how well a GAP trained only to single SIAs is able to predict the formation energies of clusters of multiple SIAs. While parallel dumbbell clusters were sufficiently well reproduced, it was not able to predict the relative stability of non-parallel SIA clusters. For example, the formation energies of the triangular \hkl<110> di-SIA and SIA clusters in the C15 Laves phase (both of which are ground-state SIA configurations in Fe, but not in W~\cite{marinica_irradiation-induced_2012}), were underestimated and therefore too stable. To correct this, we added structures containing two SIAs to the training database, including parallel and non-parallel dumbbells, and in the form of the smallest possible C15 cluster~\cite{marinica_irradiation-induced_2012,dezerald_stability_2014}. After adding di-SIAs to the training database, we found that the GAP is able to predict the energies of larger clusters in excellent agreement with DFT, as will be discussed later. All of the above-mentioned vacancy and SIA structures were sampled from MD simulations at 500--1000 K using an early version of the GAP. We note that several of the SIA structures did not remain stable during the MD preparation simulations (for example, the \hkl<110> and \hkl<100> SIAs rotate towards the \hkl<111> configuration). We included several of these unstable, rotating SIA configurations in the training database in order to capture various migration and rotation paths.

Liquid structures were added iteratively until the predicted errors of newly prepared structures were below around 10 meV/atom. The first few liquid structures were prepared in MD simulations using the existing W GAP~\cite{szlachta_accuracy_2014}. An initial GAP version trained to these structures was then used to run MD and sample additional liquid structures. We considered a range of different densities around the experimental density of liquid tungsten 17.6 g/cm$^3$~\cite{rumble_crc_2019}, including  clearly unphysical low-density liquids to ensure that the GAP does not stabilise any spurious low-density structures. We also included half-molten structures to capture the melting process.

Low-index bcc surface structures were taken from~\cite{szlachta_accuracy_2014} (\hkl(111) surfaces from \cite{dragoni_achieving_2018}). Additionally, to make our GAP applicable to surface irradiation and improve the transferability to arbitrary surface structures, we also included damaged and half-molten $\hkl(110)$ and $\hkl(100)$ surfaces. These structures were prepared by high-temperature MD simulations using an early version of the GAP that was trained to most of the remaining database, including all liquid structures and the clean surfaces.

\begin{figure}
 \centering
  \begin{subfigure}{\linewidth}
  \centering
  \includegraphics[width=\linewidth]{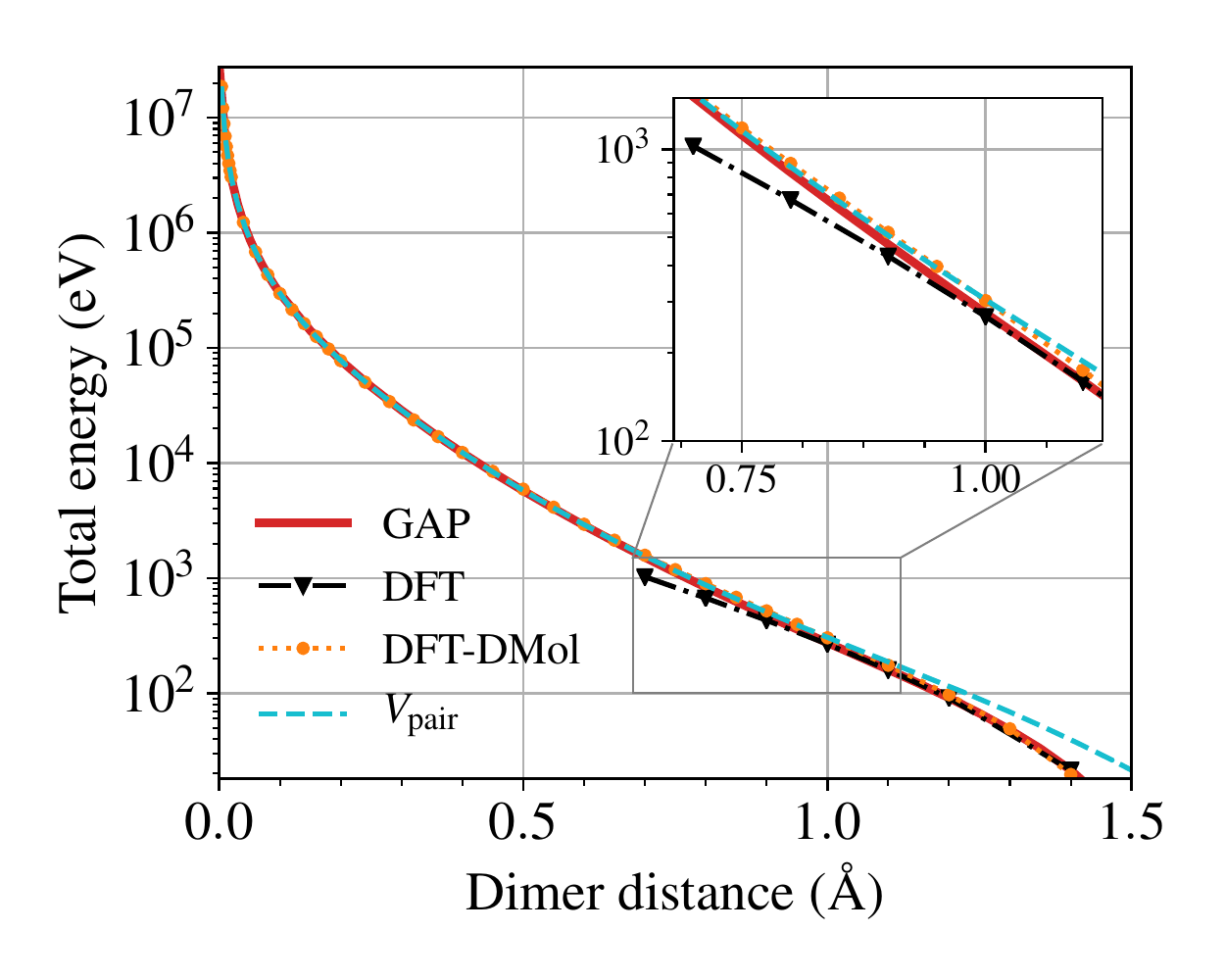}
 \end{subfigure}
 \begin{subfigure}{\linewidth}
  \centering
  \includegraphics[width=\linewidth]{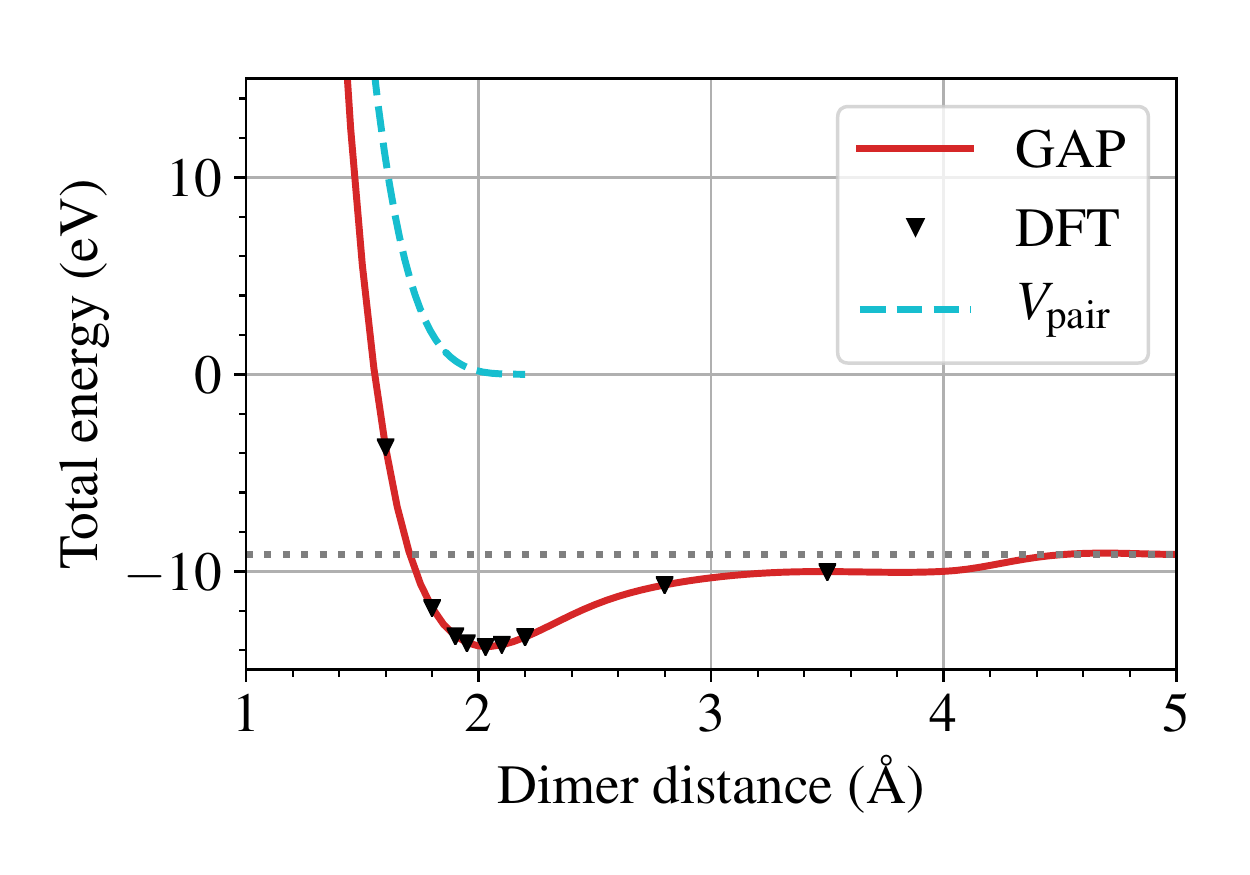}
 \end{subfigure}
 \caption{Top: short-range repulsion of a W--W dimer given by our DFT, the all-electron DFT-DMol data~\cite{nordlund_repulsive_1997}, the screened Coulomb potential $V_\mathrm{pair}$, and the trained GAP. The inset highlights the divergence of DFT compared to DFT-DMol at around 1 Å, which is used when sampling structures for the training database. Bottom: The near-equilibrium part of the dimer curve.}
 \label{fig:dimer}
\end{figure}

To ensure a physically reasonable and smooth dissociation of atoms as well as to guide the repulsive potential fit, we include energies and forces from the dimer dissociation curve. Fig.~\ref{fig:dimer} shows the dimer curve as given by our DFT calculations, compared with all-electron DFT data obtained by the DMol code~\cite{nordlund_repulsive_1997}. Our DFT results, which treats the core-electrons as frozen with the PAW formalism, are in good agreement with the all-electron DFT-DMol data down to about 1 Å, below which there is a clear divergence from DFT-DMol. Hence, we only include dimer distances larger than 1.1 Å in the training database, for which the DFT data closely overlaps with DFT-DMol and the fitted $V_\mathrm{pair}$. This ensures a smooth connection between the trained GAP and $V_\mathrm{pair}$, as the GAP is trained to predict negligible energies and forces for interatomic distances below 1.1 Å. The behaviour of the GAP at short interatomic distances is further investigated in Appendix~\ref{app:reppot}.

For capturing the short-range many-body dynamics in bcc tungsten encountered in collision cascades, we prepared various bcc crystals with a randomly added interstitial atom (called ''short-range'' in Tab.~\ref{tab:training_data}). The shortest allowed distance between the added atoms and its neighbours was 1.1 Å, corresponding to the lower limit of the range where DFT with frozen core-electrons is accurate, as discussed above. A rich variety of short-range environments was captured by adding the randomly placed atom to both perfect crystalline structures, and systems containing one or two vacancies. In part of the vacancy structures, the atom was inserted in a random position around the vacancy. Hence, in addition to sampling the non-equilibrium geometries similar to the early stages of an energetic recoil event, these structures also capture arbitrary vacancy migration paths.

The entire training database contains around 40,000 local atomic environments, which is considerably fewer than many previous single-element GAPs~\cite{szlachta_accuracy_2014,dragoni_achieving_2018,bartok_machine_2018}. Indeed, we aimed to keep the training database relatively small in anticipation of re-using the same structures for other non-magnetic bcc metals and as a basis for alloy potentials. For this reason, we decided to omit structures related to screw dislocations and gamma surfaces, which made up a large fraction of the training database for the previous tungsten GAP~\cite{szlachta_accuracy_2014} and the iron GAP~\cite{dragoni_achieving_2018}.

When training the GAP, different weights are assigned to different structure types through the regularisation parameters $\sigma_\nu$. For liquids, short-range, and the dimer structures we used $\sigma_\nu^\mathrm{energy}=10$ meV/atom, $\sigma_\nu^\mathrm{force}=0.4$ eV/Å. For disordered surfaces $\sigma_\nu^\mathrm{energy}=10$ meV/atom, $\sigma_\nu^\mathrm{force}=0.2$ eV/Å, and for all other structures $\sigma_\nu^\mathrm{energy}=1$ meV/atom, $\sigma_\nu^\mathrm{force}=0.04$ eV/Å. Virial stresses were only trained for the distorted crystal unit cells, using $\sigma_\nu^\mathrm{virial}=0.04$ eV. The resulting  root-mean-square errors (RMSE) of the training data are consistent with the assumed uncertainties, being well below 1 meV/atom and 0.1 eV/Å for most crystalline structures, and a few meV/atom and around 0.3--0.4 eV/Å for the liquid and short-range structures.

\section{Validation}
\label{sec:results}

In the following sections we present results from benchmarking of the GAP, including properties that by design are well-represented by the training database, as well as properties that were not specifically targeted in the construction of the training data. We attempt to highlight both the strengths and shortcomings of the GAP, to demonstrate the applications for which the GAP is well-suited, but also applications where an extension of the training database would be necessary. The results are compared with experimental data when possible, with DFT data from the literature when indicated as such, and with our own DFT results otherwise.

\subsection{Bulk properties}

\begin{table}
 \centering
 \caption{Basic properties of bcc tungsten: energy per atom $E_\mathrm{bcc}$, cohesive energy $E_\mathrm{coh}$, lattice constant $a$, bulk modulus $B$ and elastic constants $C_{ij}$, \hkl(110) surface energy $E_\mathrm{surf}$, vacancy formation energy $E_\mathrm{f}^\mathrm{vac}$, vacancy relaxation volume $\Omega_\mathrm{rel.}^\mathrm{vac}$, vacancy migration energy $E^\mathrm{vac}_\mathrm{mig.}$, lowest SIA formation energy $E^\mathrm{SIA}_\mathrm{f}$, SIA migration energy (main path) $E^\mathrm{SIA}_\mathrm{mig.}$, and melting temperature $T_\mathrm{melt}$.}
 \label{tab:bulk}
 \begin{threeparttable}
  \begin{tabular}{llll}
   \toprule
   & Exp. & DFT & GAP \\
   \midrule
    $E_\mathrm{bcc}$ (eV/atom) & & $-12.956$ & $-12.956$  \\
    $E_\mathrm{coh}$ (eV/atom) & $-8.81$\tnote{a} & $-8.39$ & $-8.39$ \\ 
    $a$ (Å) & 3.165\tnote{a} & 3.1854 & 3.1852 \\
    $B$ (GPa) & 310\tnote{a} & 304 & 309 \\
    $C_{11}$ (GPa) & 522\tnote{a} & 522 & 526 \\ 
    $C_{12}$ (GPa) & 204\tnote{a} & 195 & 200 \\ 
    $C_{44}$ (GPa) & 161\tnote{a} & 148 & 149 \\ 
    $E_\mathrm{surf}$ (meV/Å$^2$) & 187\tnote{b}, 203\tnote{b} & 204 & 204 \\
    $E^\mathrm{vac}_\mathrm{f}$ (eV) & $3.67\pm0.2$\tnote{c} & 3.36\tnote{d}, 3.22\tnote{e} & 3.32 \\
    $\Omega^\mathrm{vac}_\mathrm{rel.}$ & & $-0.36$\tnote{d}, $-0.33$\tnote{e} & $-0.31$ \\
    $E^\mathrm{vac}_\mathrm{mig.}$ (eV) & $1.7$--$1.9$\tnote{c} \tnote{,} \tnote{f} & 1.73\tnote{g} & 1.71 \\
    $E^\mathrm{SIA}_\mathrm{f}$ (eV) & & 10.25\tnote{h} & 10.34 \\
    $E^\mathrm{SIA}_\mathrm{mig.}$ (eV) & $<0.1$\tnote{f} & 0.040\tnote{i} & 0.038 \\
    $T_\mathrm{melt}$ (K) & 3687\tnote{a} & $3450\pm100$\tnote{j} & $3540\pm10$ \\
   \bottomrule
  \end{tabular}
  \begin{tablenotes}
  \item[a] Ref.~\cite{rumble_crc_2019}
   \item[b] Ref.~\cite{tyson_surface_1977}
   \item[c] Ref.~\cite{rasch_quenching_1980}
   \item[d] 53 atoms
   \item[e] 120 atoms
   \item[f] Ref.~\cite{heikinheimo_direct_2019}
   \item[g] Ref.~\cite{ma_effect_2019}
   \item[h] Ref.~\cite{ma_universality_2019}
   \item[i] Ref.~\cite{ma_symmetry-broken_2019}
   \item[j] Ref.~\cite{wang_melting_2011}
  \end{tablenotes}
 \end{threeparttable}
\end{table}

\begin{figure}
 \centering
 \includegraphics[width=\linewidth]{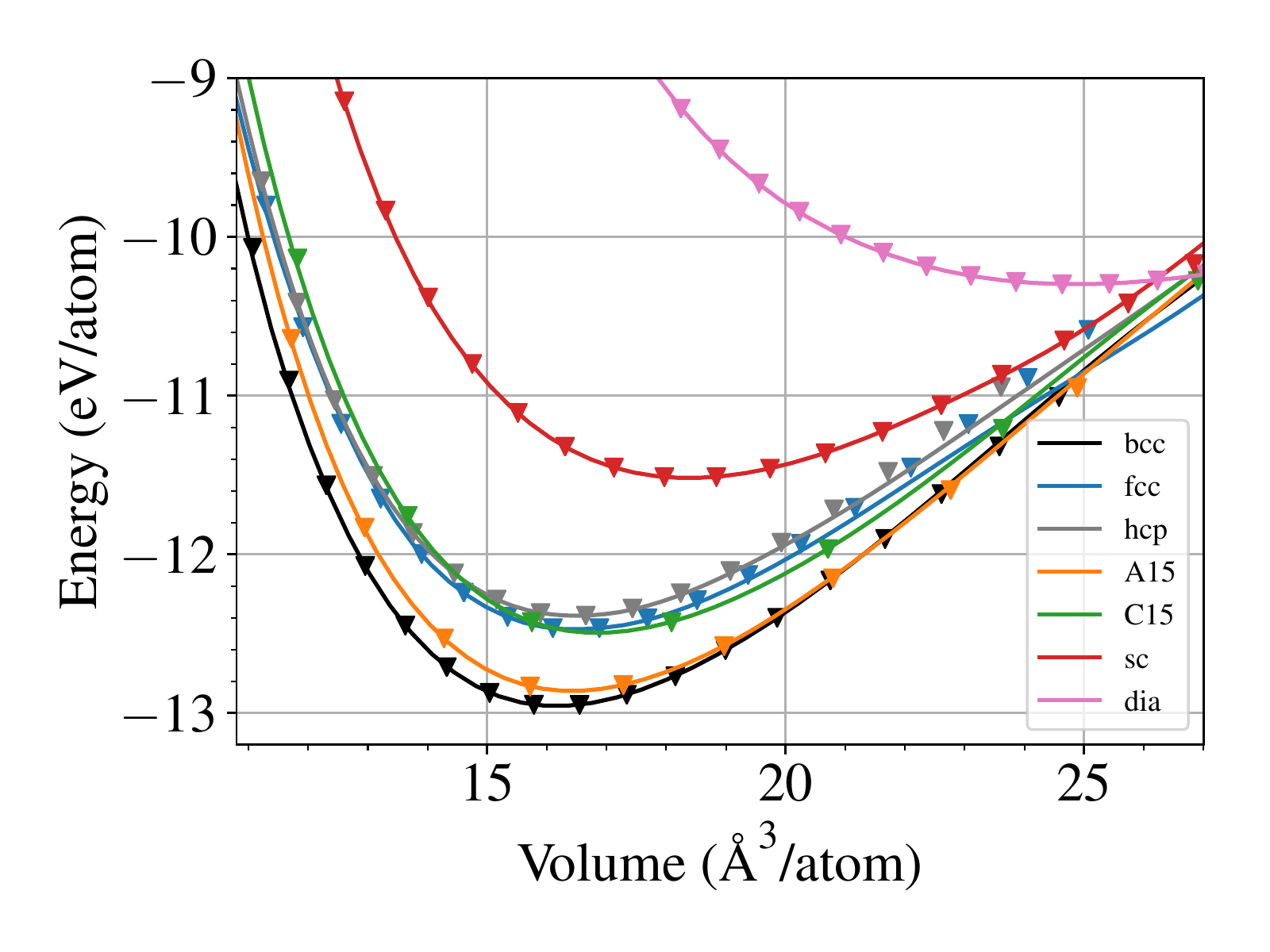}
 \caption{Energy-volume curves of the various crystal structures included in the training database. The data points are DFT data and the solid lines are the GAP predictions.}
 \label{fig:E-V}
\end{figure}

Basic properties of bcc tungsten are compiled in Tab.~\ref{tab:bulk}. All listed properties are well-represented by the training database, and therefore in close agreement with DFT. We note that the vacancy formation energy is surprisingly sensitive to the size of the box in DFT, even though elastic interactions across the periodic boundaries are negligible~\cite{ma_universality_2019} and the $k$-point density is the same. With a box of 53 atoms we obtain a formation energy of 3.36 eV, while a box of 120 atoms gives 3.22 eV. Almost identical values have been reported previously~\cite{ma_universality_2019,ma_effect_2019} (3.35 for a 53-atom box and 3.22 for a 127-atom box). Since our training database contains structures of different sizes, the GAP reproduces a value in-between these two values (regardless of box size).

Fig.~\ref{fig:E-V} shows energy-volume curves of various crystal structures. All of these crystals were included in the training database (although only as randomly distorted cells) and the GAP therefore accurately reproduce the DFT data. The only visible discrepancies are for strongly expanded fcc and hcp lattices ($> 20$ Å$^3$/atom).

\begin{figure}
 \centering
 \begin{subfigure}{\linewidth}
  \centering
  \includegraphics[width=0.9\linewidth]{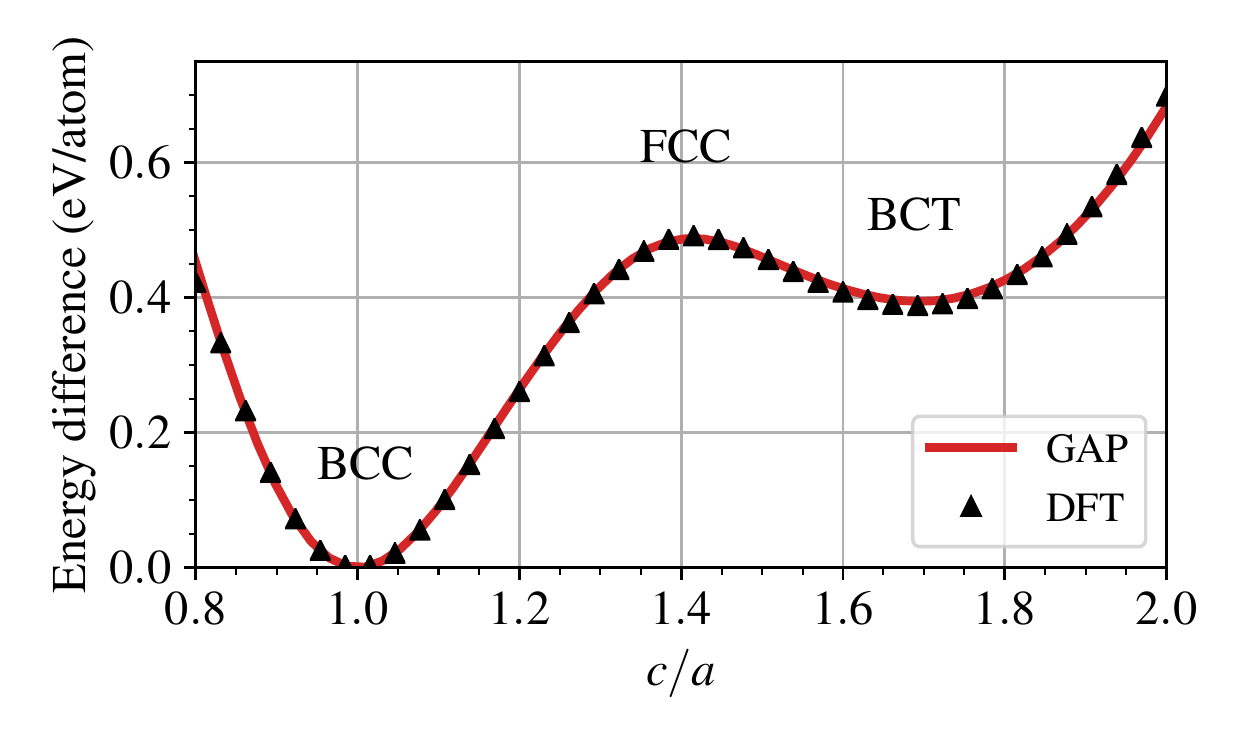}
  \caption{Tetragonal (Bain) path.}
 \end{subfigure}
  \begin{subfigure}{\linewidth}
  \centering
  \includegraphics[width=0.9\linewidth]{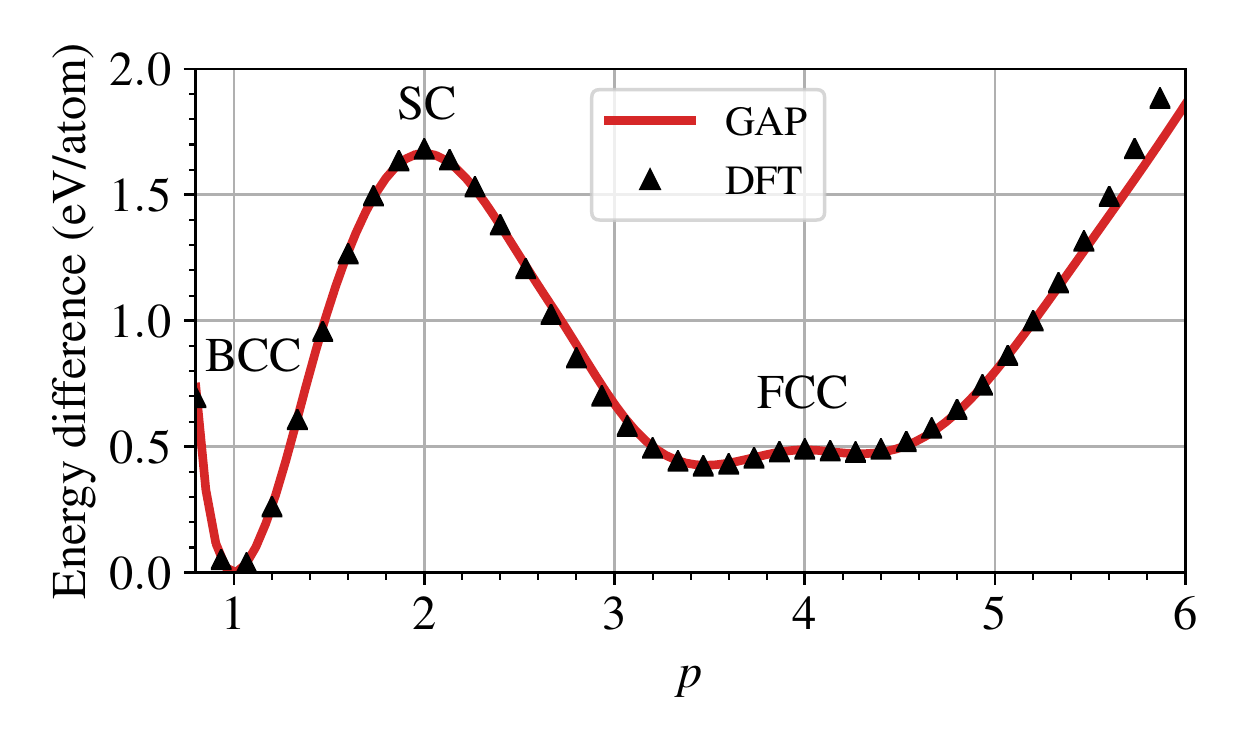}
  \caption{Trigonal path.}
 \end{subfigure}
  \begin{subfigure}{\linewidth}
  \centering
  \includegraphics[width=0.9\linewidth]{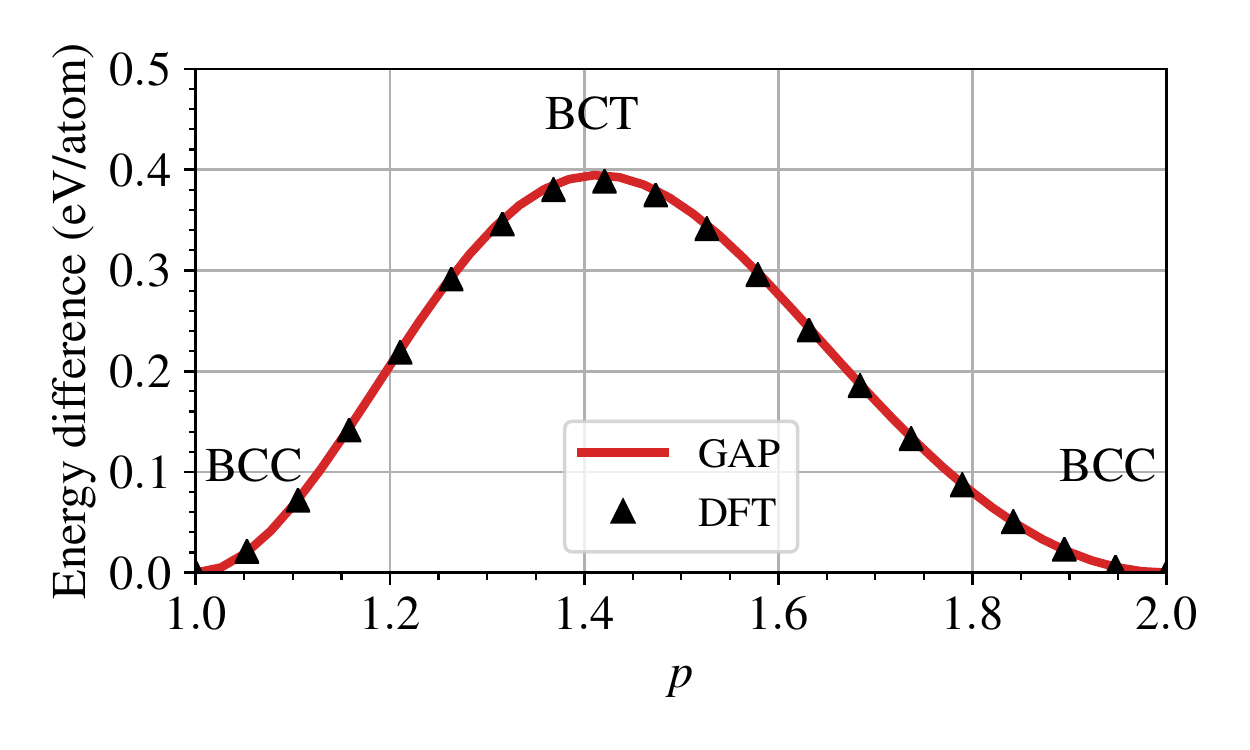}
  \caption{Orthorhombic path.}
 \end{subfigure}
  \begin{subfigure}{\linewidth}
  \centering
  \includegraphics[width=0.9\linewidth]{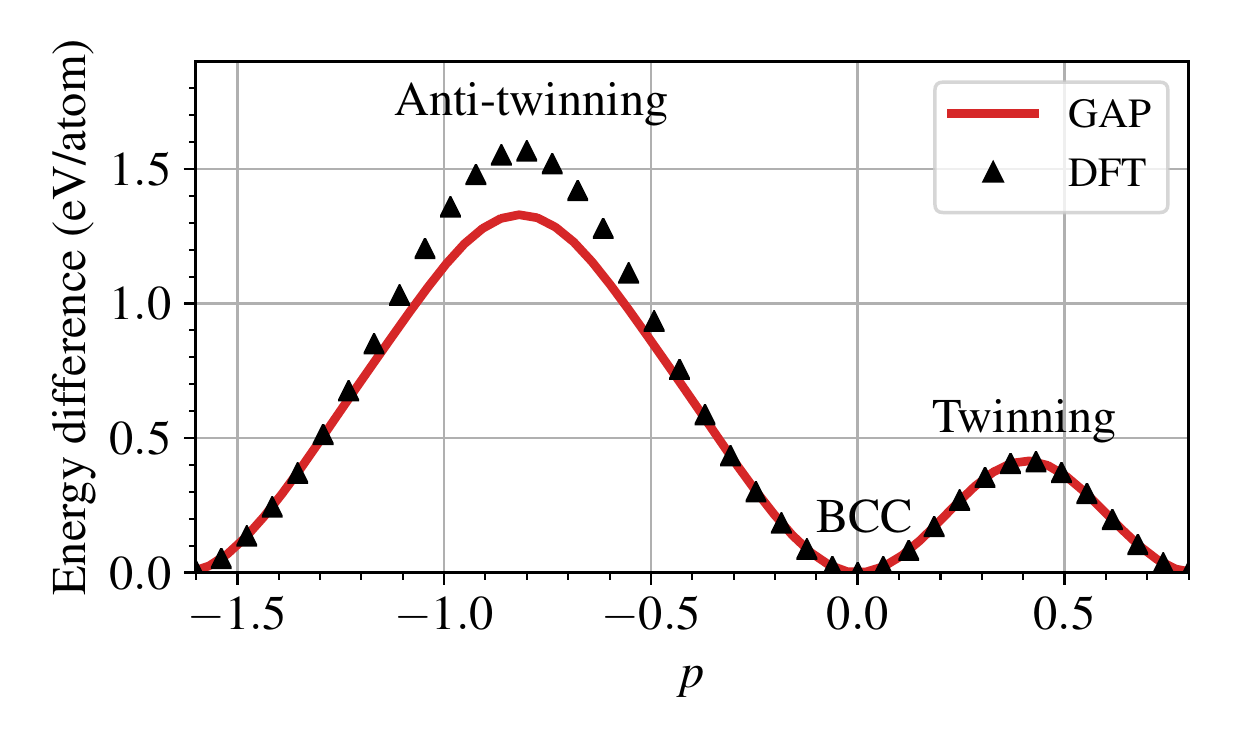}
  \caption{Twinning and anti-twinning of bcc.}
 \end{subfigure}
 \caption{Volume-conserving deformation paths of bulk W computed with GAP and DFT.}
 \label{fig:bulk_deform}
\end{figure}

To further explore the transferability of the GAP to crystal symmetries not included in the training database, we considered four different volume-conserving deformation paths of the bcc crystal. The tetragonal path (also called the Bain path) is perhaps the most well-known path, in which a bcc crystal is stretched along the \hkl[100] direction and simultaneously compressed in \hkl[010] and \hkl[001], leading to the fcc symmetry and eventually the body-centred tetragonal (bct) crystal. For the trigonal path, the bcc crystal is stretched along the \hkl[111] direction and compressed in \hkl[1-10] and \hkl[11-2], reaching the simple-cubic symmetry and eventually fcc again. The orthorhombic deformation path involves stretching in the \hkl[001] direction while compressing in \hkl[110]. Finally, twinning and anti-twinning involves shearing a primitive bcc lattice in \hkl[-1-11] (positive strains for twinning and negative strains for anti-twinning) and can be used to measure the theoretical shear strength of single crystals~\cite{paxton_quantum_1991}. The energy difference for each of these deformation paths are shown in Fig.~\ref{fig:bulk_deform}, where the GAP results are compared with our DFT data. The values of $c/a$ and $p$ correspond to the magnitude of the strains. For more details about the various deformation paths, we refer to Refs.~\cite{paidar_study_1999,mrovec_bond-order_2007,ravelo_shock-induced_2013}. For the most part, GAP is indistinguishable from DFT, with the only notable exceptions being underestimating the anti-twinning energy and the high-strain tail of the trigonal path. Note that all deformation paths correspond to strains far beyond the maximum strains of the training structures.

\begin{figure}
 \centering
 \includegraphics[width=\linewidth]{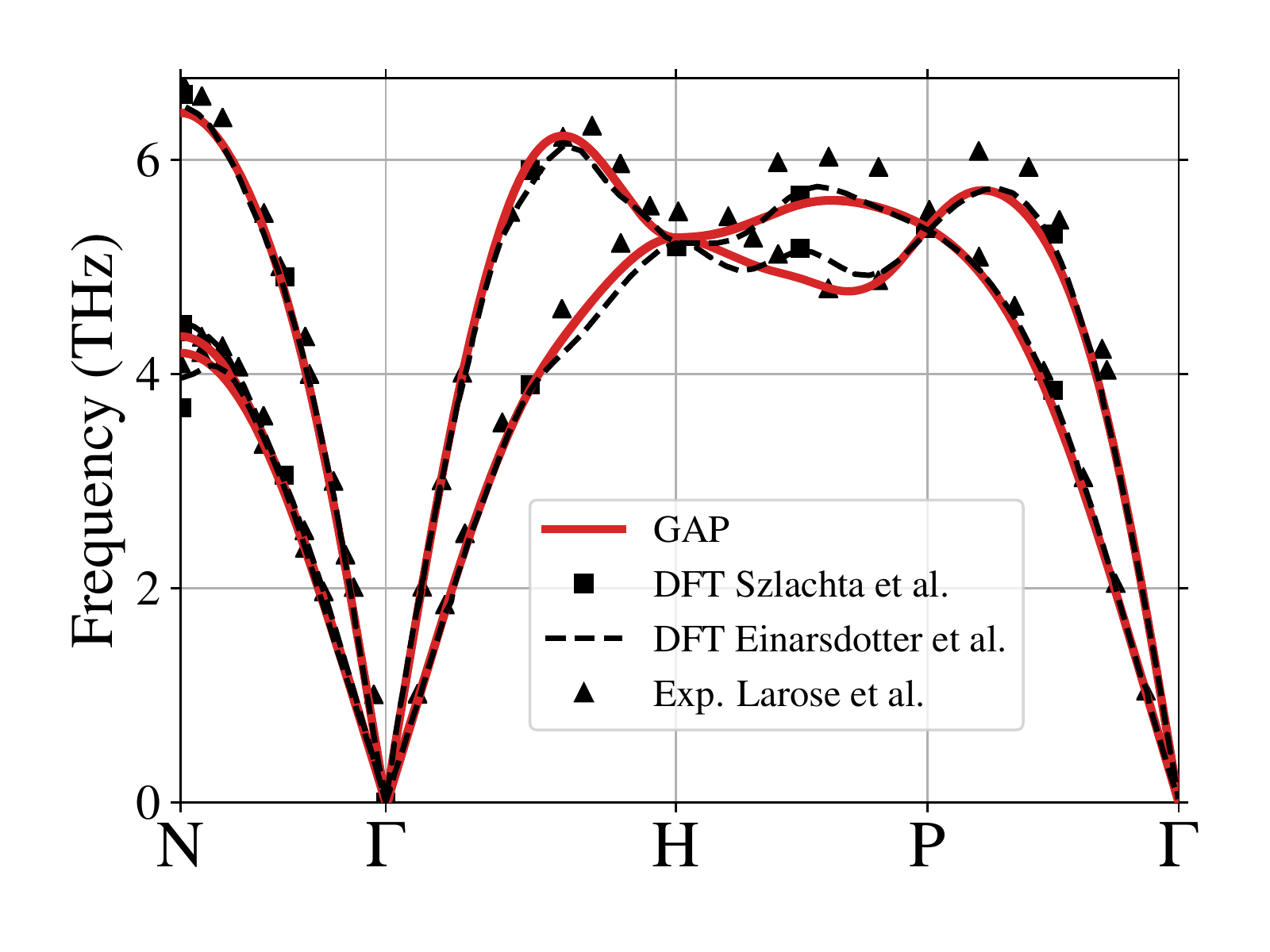}
 \caption{Phonon dispersion of bcc W as given by the GAP and compared with DFT~\cite{einarsdotter_phonon_1997,szlachta_accuracy_2014} and experimental data~\cite{larose_lattice_1976}.}
 \label{fig:phonon}
\end{figure}

Fig.~\ref{fig:phonon} shows the phonon dispersion of bcc tungsten compared with experimental data and previous DFT studies~\cite{einarsdotter_phonon_1997,szlachta_accuracy_2014,larose_lattice_1976}. The dispersion relation is overall well-reproduced by the GAP, although subtle discrepancies exist, in particular between the H and P points and at the N point. It remains unclear what causes these differences between GAP and DFT, which were also observed in the previous tungsten GAP~\cite{szlachta_accuracy_2014} (the phonon dispersions in both GAPs are virtually identical).

\begin{figure}
 \centering
 \includegraphics[width=\linewidth]{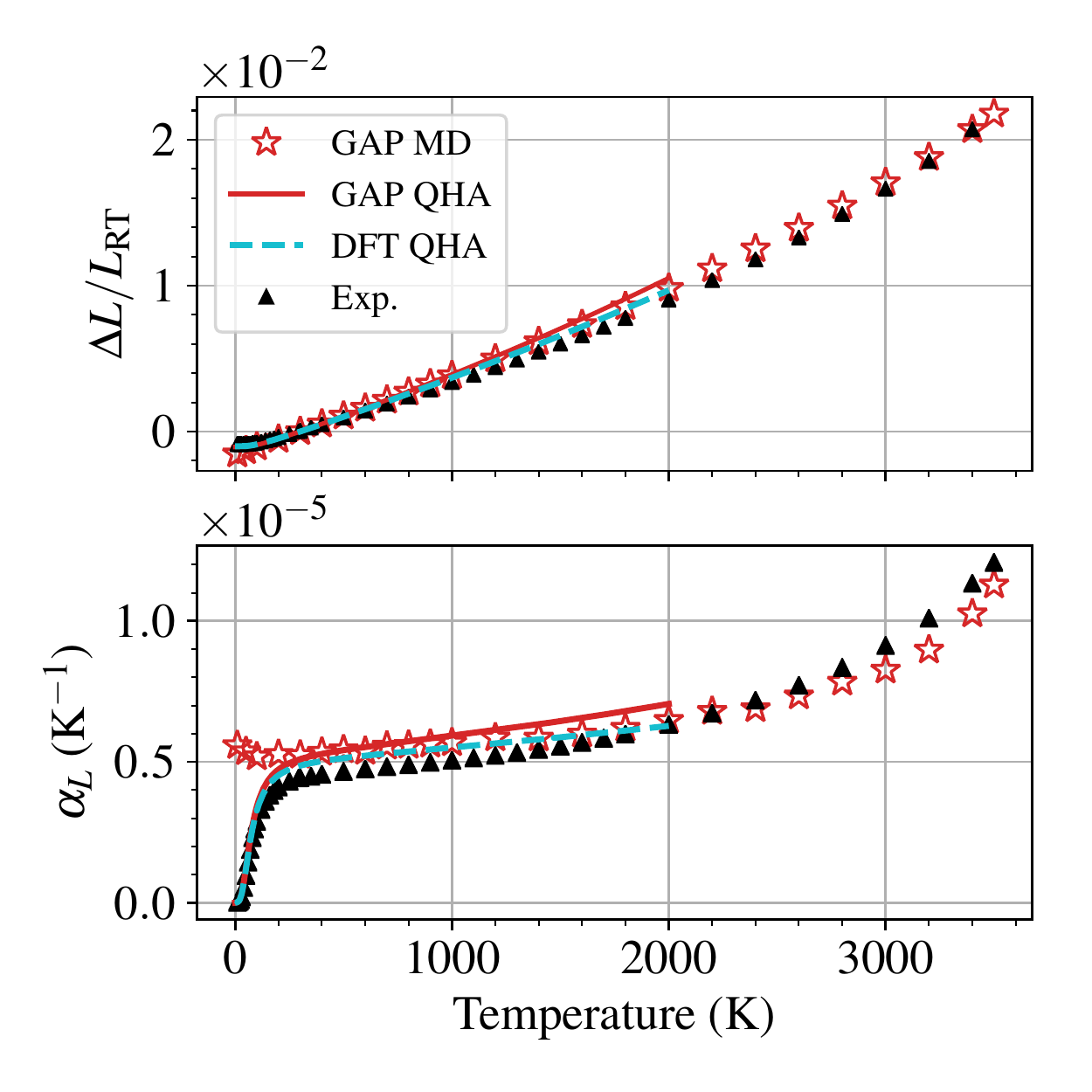}
 \caption{Linear thermal expansion (top) and the expansion coefficient (bottom) of bcc W predicted by the GAP and compared with experimental results~\cite{white_thermophysical_1997}. GAP data is obtained from both MD simulations and by using the quasi-harmonic approximation (QHA).}
 \label{fig:thermexp}
\end{figure}

\begin{table}
 \centering
 \caption{Heat capacities ($C_P$, $C_V$), linear thermal expansion coefficient ($\alpha_L$), and Gr\"uneisen parameter ($\gamma$) at 300 K calculated within the quasi-harmonic approximation with GAP and DFT, and compared with experimental data.}
 \label{tab:therm_prop}
 \begin{threeparttable}
  \begin{tabular}{llll}
   \toprule
   & Exp. & DFT & GAP \\
   \midrule
   $C_P$ (J mol$^{-1}$K$^{-1}$) & 24.35\tnote{a} & 23.95 & 23.98 \\
   $C_V$ (J mol$^{-1}$K$^{-1}$) & 24.20\tnote{a} & 23.77 & 23.77 \\
   $\alpha_L$ (10$^{-6}$K$^{-1}$) & 4.43\tnote{b} & 4.87 & 5.10 \\
   $\gamma$ & 1.6\tnote{a} & 1.80 & 1.87 \\
   \bottomrule
  \end{tabular}
 \begin{tablenotes}
   \item[a] Ref.~\cite{white_heat_1984}
   \item[b] Ref.~\cite{white_thermophysical_1997}
  \end{tablenotes}
 \end{threeparttable}
\end{table}

The linear thermal expansion and the associated expansion coefficient ($\alpha_L$) of bcc tungsten as predicted by the GAP is shown in Fig.~\ref{fig:thermexp}, and compared with experimental measurements from~\cite{white_thermophysical_1997} and our DFT results. The expansion is calculated with the reference temperature set to room temperature (300 K), as in the experimental data. GAP data is obtained by two different methods; MD simulations in the $NPT$ ensemble, and calculations within the quasi-harmonic approximation (QHA) using the \textsc{phonopy} code~\cite{togo_first_2015}. The latter includes zero-point energies and is accurate at low temperatures, but eventually fails when anharmonic effects beyond the QHA become non-negligible. On the other hand, MD fails at low temperatures but is reliable at temperatures when zero-point energies are negligible. Fig.~\ref{fig:thermexp} shows that the experimentally measured low-temperature trend is well-captured by both GAP and DFT when combined with the QHA. Fig.~\ref{fig:thermexp} also suggests that the QHA is valid up to around 1000 K, while MD with the GAP is consistent with the experimental trend at temperatures above 300 K. The thermal expansion coefficients at room temperature are listed in Tab.~\ref{tab:therm_prop}, along with heat capacities and the Gr\"uneisen parameter. The experimental heat capacity is well-reproduced by both GAP and DFT. DFT overestimates the experimental room-temperature thermal expansion coefficient and Gr\"uneisen parameter by about 10\%, and GAP by about 15\%.

\begin{figure}
 \centering
 \includegraphics[width=\linewidth]{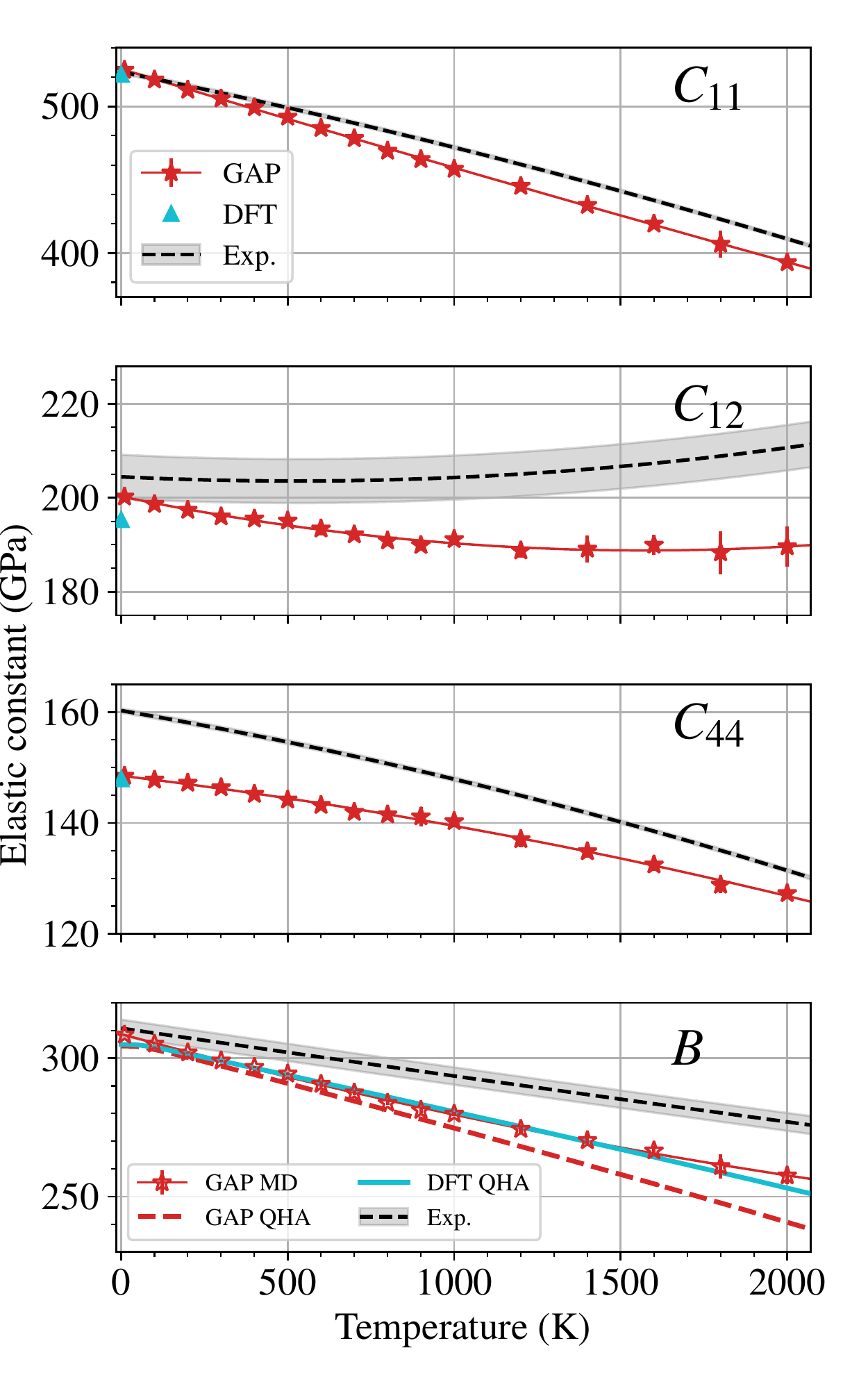}
 \caption{Elastic constants and bulk modulus of bcc W as functions of temperature. Experimental data are from Ref.~\cite{lowrie_singlecrystal_1967}. GAP data for the bulk modulus is obtained from both MD simulations and by using the quasi-harmonic approximation (QHA). DFT data are shown at 0 K for the elastic constants and as obtained by the QHA for the bulk modulus. The solid lines connecting the GAP points are polynomial fits to guide the eye.}
 \label{fig:eltemp}
\end{figure}

Fig.~\ref{fig:eltemp} shows the elastic constants of bcc W at finite temperatures as predicted by the GAP. The results are compared to the experimental least-squares fits from Ref.~\cite{lowrie_singlecrystal_1967}, measured for single crystals up to around 2000 K. The uncertainties of the experimental curves are shown as shaded areas. The GAP elastic constants are extracted from the average stress tensor of constant-temperature MD simulations of distorted bcc systems containing 1024 atoms. The error bars are given by the standard deviation of the values obtained for equivalent elastic constants (e.g. $C_{11}$, $C_{22}$, and $C_{33}$). The experimental trends are qualitatively well reproduced by the GAP, although quantitative differences are apparent. Both $C_{11}$ and $C_{44}$ decrease at increasing temperatures, while $C_{12}$ remains almost constant at low temperatures and increases slightly at higher temperatures. The weak temperature dependence of $C_{12}$ is reproduced by the GAP, although the uncertainties in both experiments and MD are larger than for the other elastic constants. The softening of the bulk modulus in DFT can be estimated from finite-temperature free energy-volume curves calculated in the quasi-harmonic approximation. Results for both DFT and the GAP coupled with the QHA are shown for the bulk modulus in Fig.~\ref{fig:eltemp}. Note that as previously mentioned, the QHA is only reliable up to around 1000 K. The good agreement between GAP and DFT for the bulk modulus indicate that the quantitative discrepancies between experiments and GAP are mainly inherited from DFT, as both GAP and DFT predict a slightly stronger temperature dependence of the bulk modulus than experiments.

 For validating that we sampled enough liquid structures, we performed a form of $k$-fold cross validation (with $k=5$). That is, we split the 45 liquid structures into five subsamples, with each part containing liquids with different densities. Five different GAP models were then trained using four of the subsamples together with the rest of the training database, while for each model leaving out a different liquid subsample for validation. The mean root-mean-square errors for the energy and forces of the validation subsamples for the five GAP models are 7.76 meV/atom and 0.434 eV/Å. These values are close to the assumed uncertainties, $\sigma_\nu$, used when training the GAP (10 meV/atom and 0.4 eV/Å). This provides confidence that the GAP reproduces the energies and forces of any liquid structure with sufficient accuracy.
 
We simulated the melting temperature predicted by the GAP using the solid-liquid interface method. At 3540 K, the solid and liquid phases remains roughly in equilibrium, while at 3550 K the entire system melts and at lower temperature it recrystallises. The estimated melting temperature of 3540 K is slightly lower, but very close to the experimental value of 3687 K~\cite{rumble_crc_2019}. Wang et al.~\cite{wang_melting_2011} used \textsc{vasp} with comparable settings to our training data (PBE functional and hard PAW potential), and estimated a melting point of $3450 \pm 100$ K using two different methods. This is in line with the GAP prediction of 3540 K, which confirms that the GAP reproduces melting with DFT accuracy. Hence, the slight underestimation compared to experiments can be attributed to the accuracy of DFT with the PBE functional.

 \subsection{Surface properties}

\begin{figure}
    \centering
    \includegraphics[width=\linewidth]{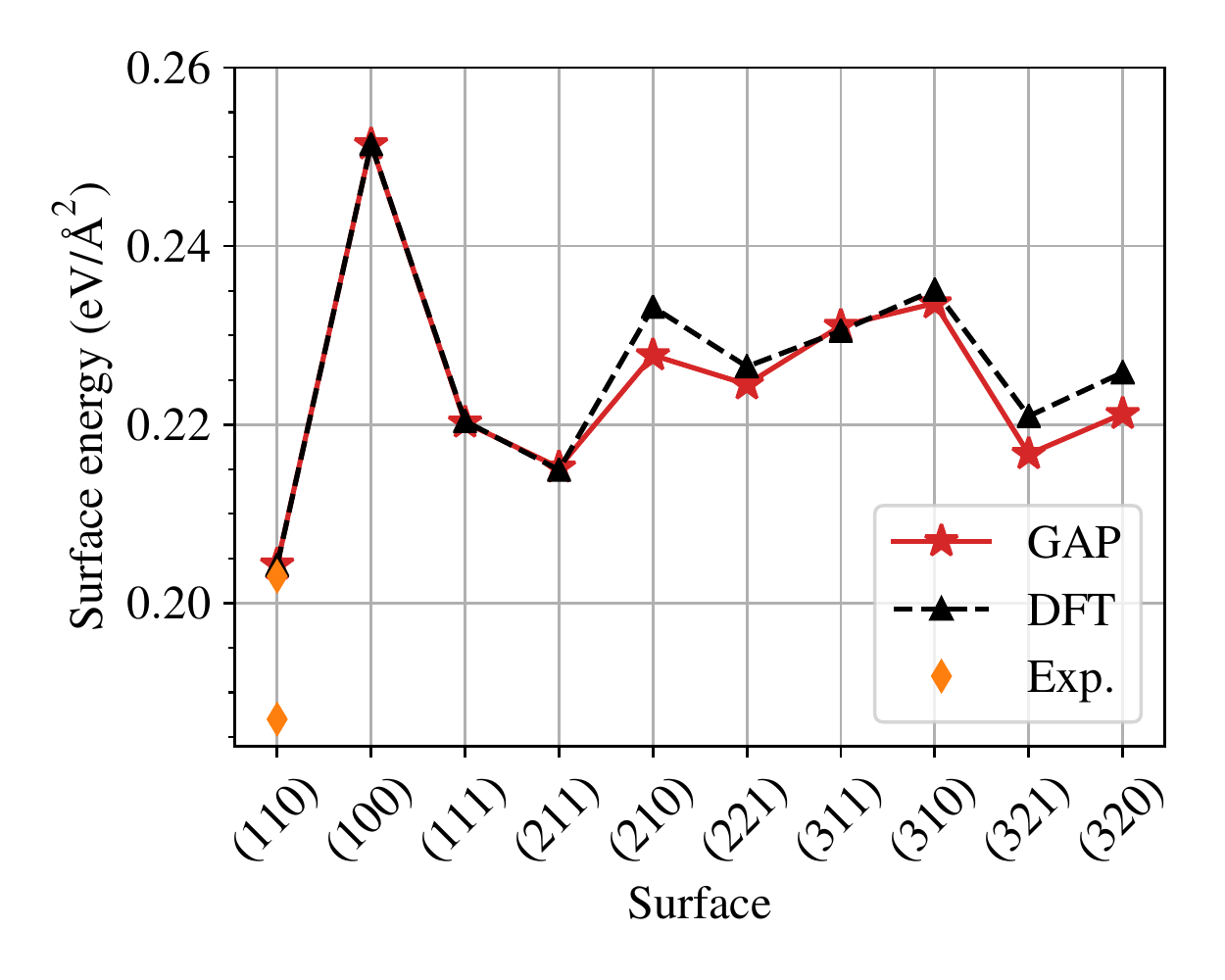}
    \caption{Surface energies predicted by the GAP and compared with DFT. Only the first four surfaces were included in the training database. The experimental values are from Ref.~\cite{tyson_surface_1977}.}
    \label{fig:esurf}
\end{figure}

We calculated surface energies of 10 surfaces with DFT in order to test how transferable the GAP is to surfaces not included in the training database. A comparison between GAP and DFT is shown in Fig.~\ref{fig:esurf}, where only the first four low-index surfaces were included in the training database. The GAP successfully predicts surface energies in close agreement with DFT, with the largest discrepancies within 5 meV/Å$^2$ of the DFT values. The order of stability is also for the most part reproduced by the GAP, although e.g. the \hkl(321) surface is incorrectly lower in energy than the \hkl(111) surface. The ability to reproduce accurate surface energies is a clear improvement over traditional analytical potentials, which consistently underestimate surface energies and fail to reproduce the correct order of stability~\cite{bonny_many-body_2014}.

We also confirmed that the GAP reproduces the DFT-observed displacements along the surface normal during relaxation of the most common surfaces. Relaxation of the most stable \hkl(110) surface involves a small shift ($-0.07$ Å) of the topmost layer down towards the bulk. For the \hkl(111) surface in both the GAP and DFT, the topmost atomic layer is relaxed by $-0.27$ Å, the second layer by $-0.08$ Å and the third layer by 0.14 Å compared to the initial bulk lattice spacing. The \hkl(112) surface undergoes a subtle spontaneous reconstruction. In both DFT and the GAP, the topmost layer is laterally displaced by 0.1 Å in the \hkl[111] direction during relaxation (0.06 Å with respect to the second surface layer and 0.11 Å with respect to the third). This is in excellent agreement with the experimentally observed lateral \hkl[111] shift of 0.1 Å~\cite{grizzi_time--flight_1989}.

\begin{figure}
 \centering
 \includegraphics[width=0.48\linewidth]{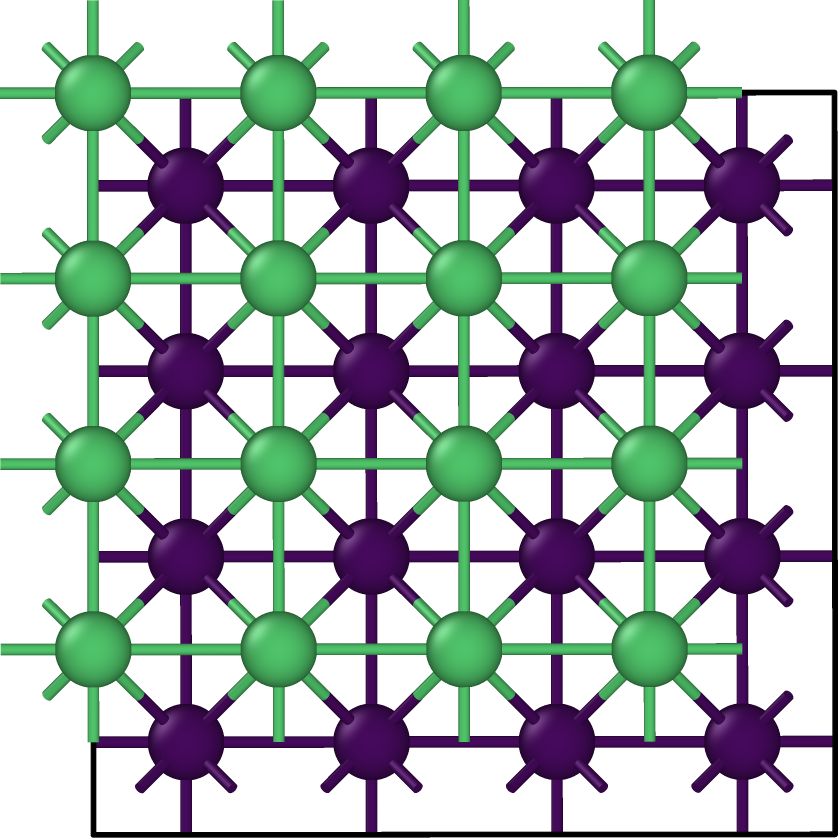}
 \includegraphics[width=0.48\linewidth]{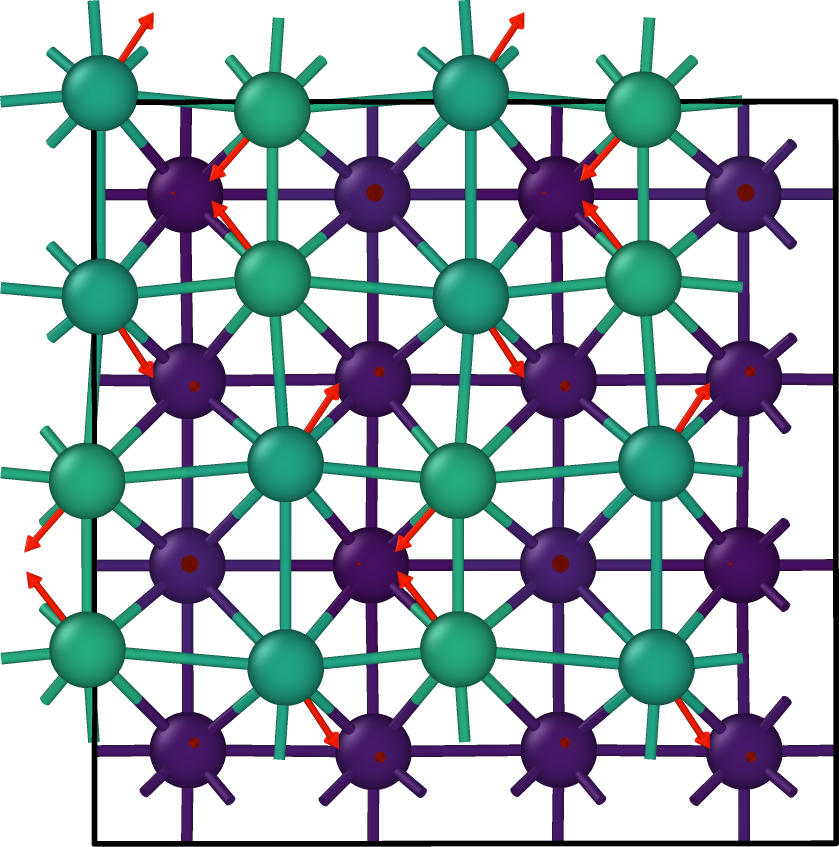}
 
 \vspace{5mm}
 
 \includegraphics[width=0.5\linewidth]{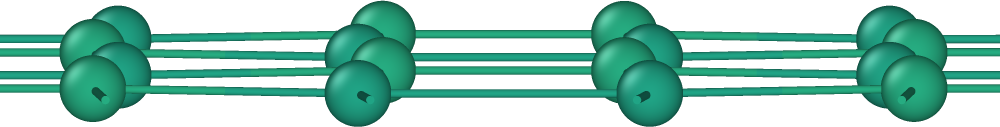}
 \caption{Reconstruction of the \hkl(100) surface. Top left: initial unrelaxed surface, top right: reconstructed surface. Only the top two surface layers are shown and atoms are coloured according to height, so that atoms in the top layer are green and atoms in the second layer purple. Red arrows show the direction of the displacement vectors with respect to the unrelaxed surface. The bottom figure shows a side-view of the zigzag surface layer.}
 \label{fig:100reconstr}
\end{figure}

For the \hkl(100) surface, there is an experimentally and theoretically observed reconstruction, in which the atoms in the surface layer are shifted laterally by a small distance, resulting in zigzag rows of atoms~\cite{debe_space-group_1977,altman_multilayer_1988}. This reconstruction does not occur spontaneously in either DFT or the GAP when optimising the atom positions, and is not explicitly included in the training database. Nevertheless, since our GAP is trained to various disordered or half-molten surface structures (the purpose of which is to at least qualitatively capture arbitrary surface properties), it is a good test to investigate whether it is able to reproduce the \hkl(100) reconstruction. Indeed, upon heating and quenching a \hkl(100) surface in MD simulations with the GAP, the \hkl(100) surface reconstructs in the way described above. Fig.~\ref{fig:100reconstr} shows snapshots of the initial and reconstructed surfaces. The lateral displacement of the surface atoms in the reconstructed layer is about 0.2 Å in the \hkl<11> direction in the surface plane, which is slightly below the DFT-obtained 0.28 Å~\cite{heinola_first-principles_2010} but coincidentally in better agreement with the experimental value 0.24 Å~\cite{altman_multilayer_1988}. The surface energy of the relaxed reconstructed surface is about 3 meV/Å$^2$ lower than for the perfect \hkl(100) surface.

In order to further test the GAP for surface properties, we carried out NEB calculations for the main migration paths of adatoms on the \hkl(110) and \hkl(100) surfaces (adatom-hopping between adjacent ground state adsorption sites, and the exchange migration as discussed in e.g.~\cite{chen_biaxial_2013}). The tests revealed that while the GAP reproduces the correct stable adsorption sites, the migration paths are systematically underestimated by about 20--30\% compared to the DFT values from~\cite{chen_biaxial_2013}. For example, the main migration mechanism of adatoms on the \hkl(110) (hopping between adjacent long-bridge sites) has a barrier of 0.87 eV according to DFT~\cite{chen_biaxial_2013}, while GAP predicts a barrier of 0.6 eV. For the exchange mechanism, the comparison is 3.09 eV by DFT and 2.5 eV by GAP. Hence, the GAP does reproduce the correct adatom behaviour in terms of stable sites and migration mechanisms, but would require an extension of the training database in order to achieve quantitative agreement with DFT (by e.g. adding various stable and unstable adatom structures to the training data). Nevertheless, we conclude that including the disordered surface structures in the training database achieved our goal of qualitatively capturing the correct behaviour of damaged surfaces.

\subsection{Repulsive potential}

\begin{figure*}
 \centering
 \includegraphics[width=0.49\linewidth]{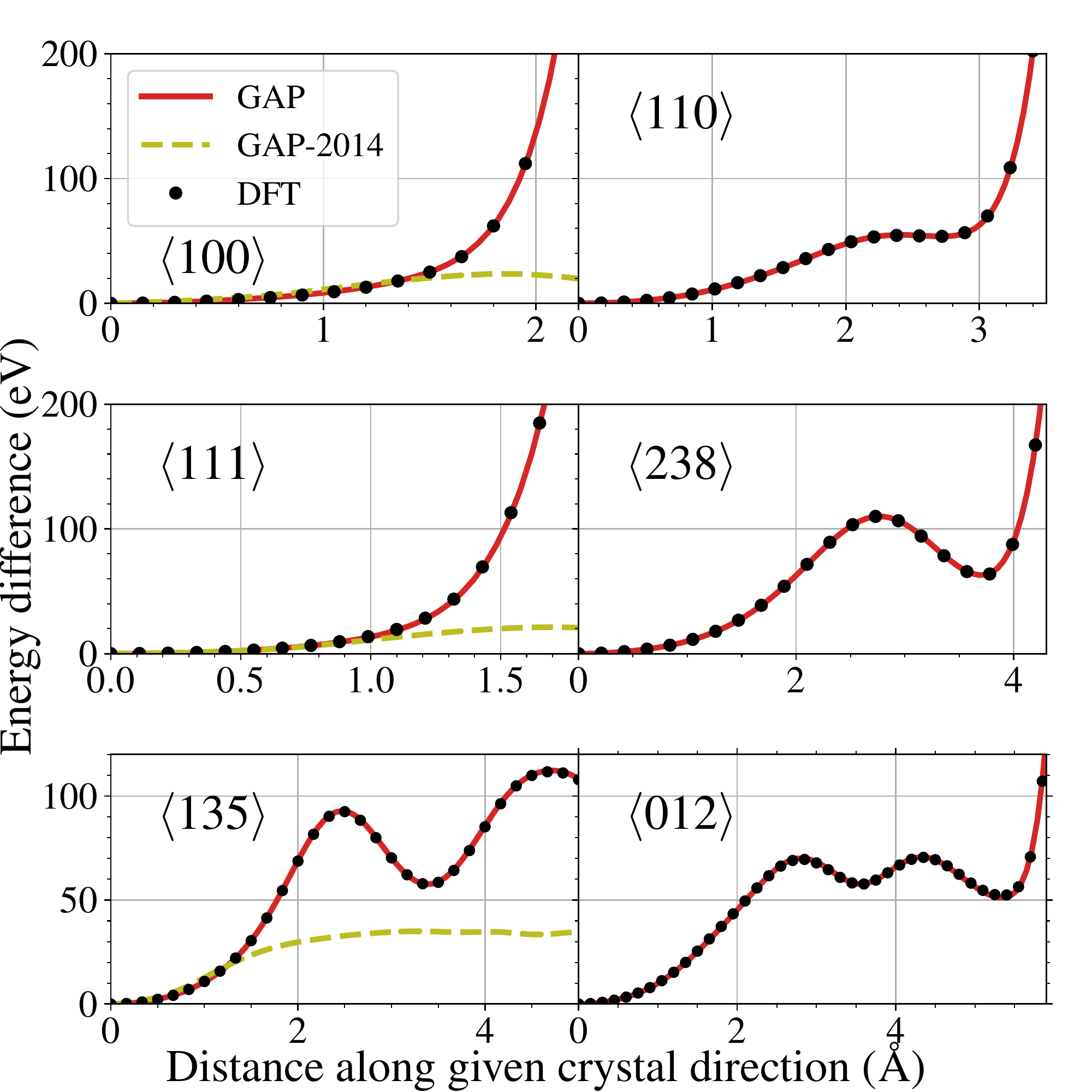}
 \includegraphics[width=0.49\linewidth]{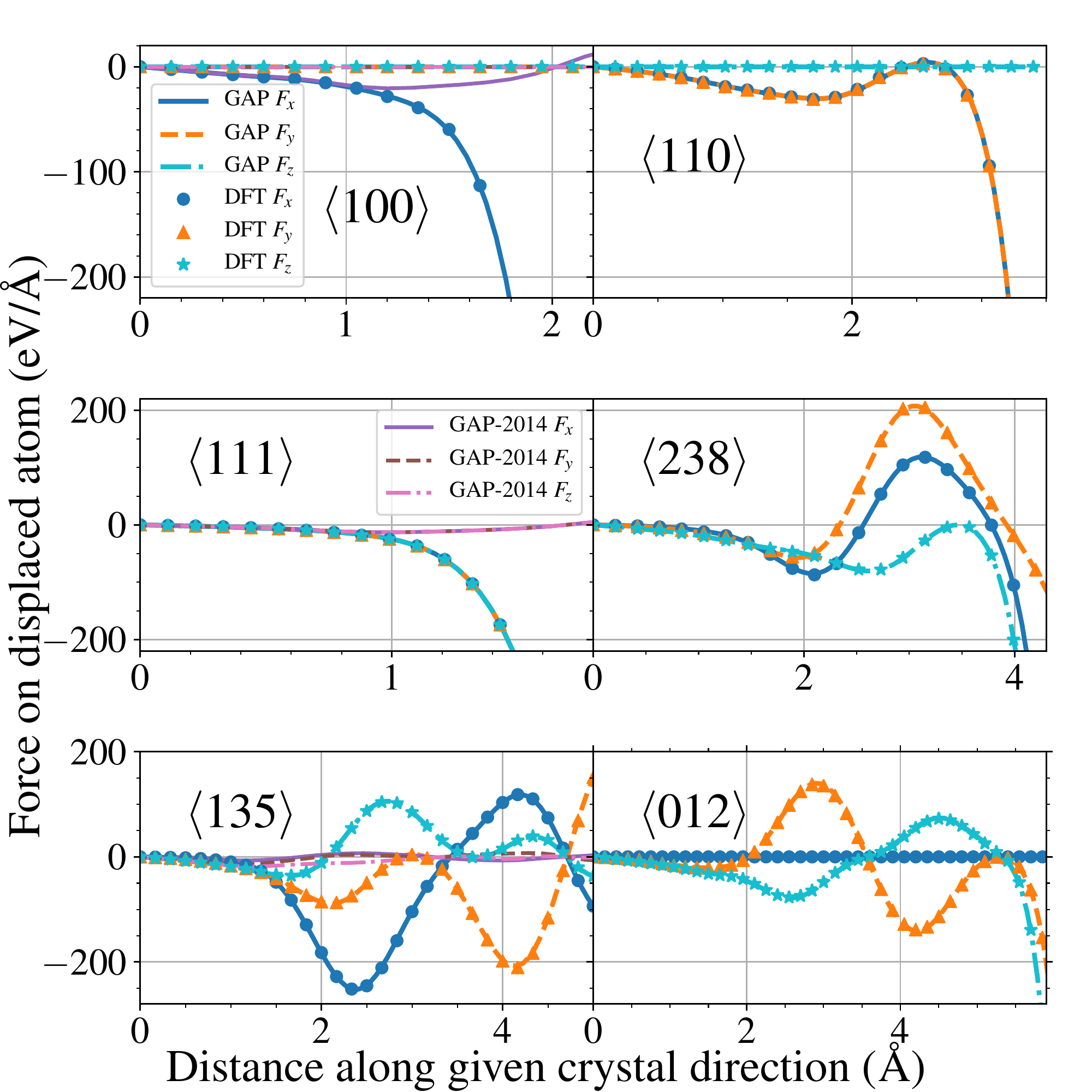}
 \caption{Total energy difference (left) and force components (right) for step-wise movement of one atom along various crystal directions in bcc W. None of the structures along these specific paths were included in the training database. Results using a previous GAP~\cite{szlachta_accuracy_2014}, not trained to any repulsive data, are shown for a few crystal directions for comparison.}
 \label{fig:qsd}
\end{figure*}

The short-range many-body behaviour relevant for cascade simulations was tested by statically moving an atom along various crystal directions in the bcc lattice. The difference in total energy and the force components of the moving atom were calculated in both GAP and DFT. Only the interatomic range for which DFT is still accurate, as discussed previously, was sampled (down to about 1.1 Å or 100--200 eV energy differences). Fig.~\ref{fig:qsd} shows the obtained curves for six different crystal directions. Several more directions were sampled with similar results, and therefore not shown here. The agreement between GAP and DFT is excellent. Considering that none of the points shown in Fig.~\ref{fig:qsd} were included in the training database, we are confident that the GAP reproduces any short-range forces and energies encountered in cascade simulations with DFT accuracy. Fig.~\ref{fig:qsd} also includes results using the previous tungsten GAP~\cite{szlachta_accuracy_2014}, demonstrating the poor extrapolation of GAP when repulsive interactions are not considered during training.

\begin{figure}
    \centering
    \includegraphics[width=\linewidth]{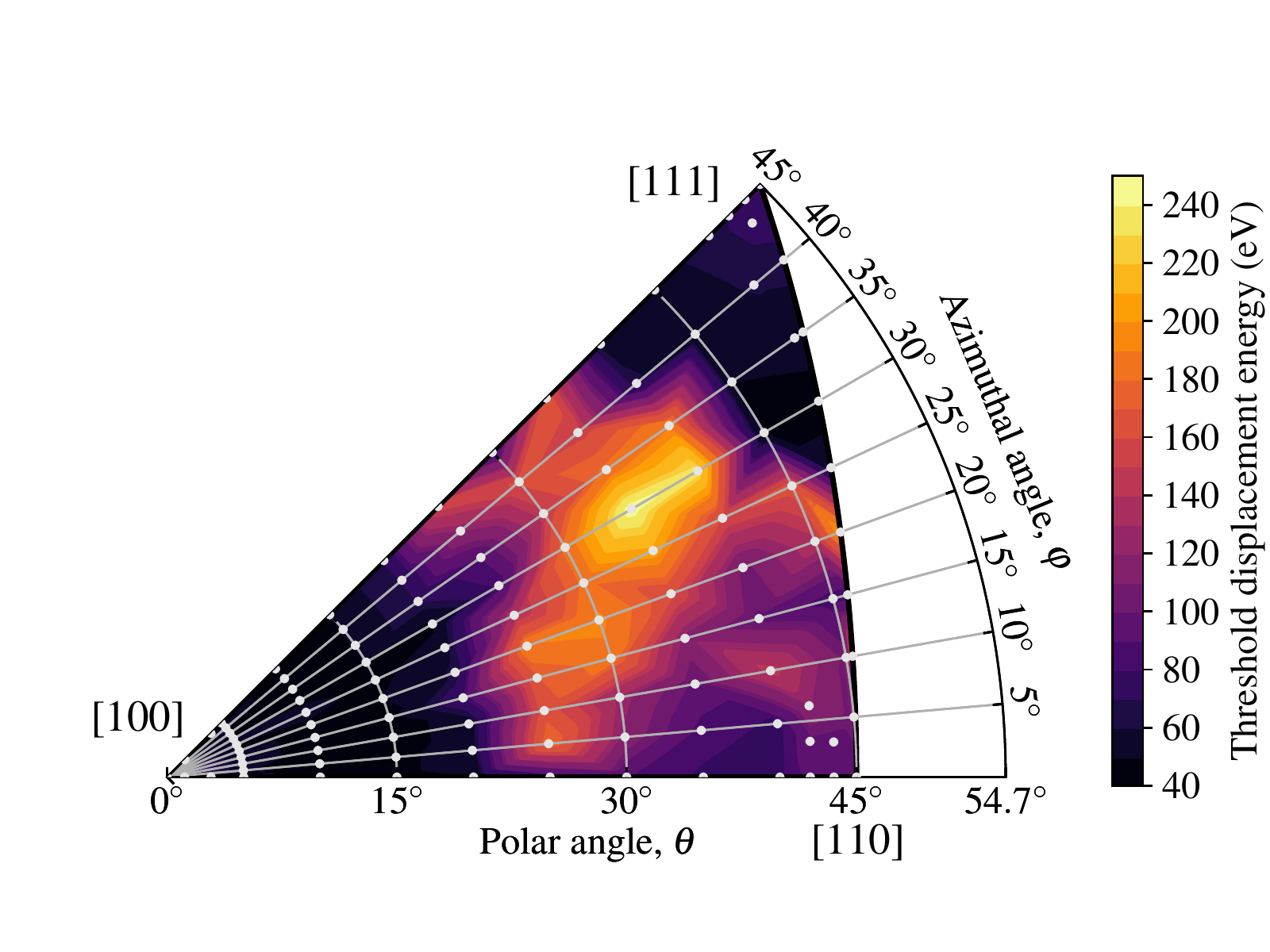}
    \caption{Threshold displacement energies obtained with the GAP at 0 K. The colours are linearly interpolated between the (light grey) data points. The average value of the uniformly sampled points is $93.6\pm5.5$ eV and the minimum values 45.5 eV for the \hkl<100> direction, 51.5 eV for \hkl<111> and 78 eV for \hkl<110>, compared with the experimental values $42\pm1$ eV for \hkl<100> and $44\pm1$ eV for \hkl<111>~\cite{maury_frenkel_1978}.}
    \label{fig:tde}
\end{figure}

To test the short-range part of the GAP in dynamic simulations, we simulated the threshold displacement energy (TDE) surface according to the methods described in Ref.~\cite{nordlund_molecular_2006}. The simulations were performed at 0 K. We also simulated a few directions in a sample equilibrated at 40 K for comparison, but found that the minimum values for a given direction remained the same as for 0 K. Hence, we report the results obtained at 0 K, for which we can exploit the full symmetry of the lattice when sampling directions. The crystal directions were sampled uniformly over the symmetry-reduced sphere at 5$\degree$ intervals. We used a non-cubic simulation box of 4368 atoms. The increment in kinetic energy was 4 eV. After obtaining the full angular map of TDEs, we sampled additional directions close to the low-index directions with a lower increment of 1 eV to obtain more exact TDE values for comparison with experiments.

Fig.~\ref{fig:tde} shows the angular map of the threshold displacement energies obtained at 0 K with the GAP. The global average of the uniformly sampled directions is $93.6\pm5.5$ eV. As expected based on experimental results~\cite{maury_frenkel_1978}, the minimum TDE values are found around the \hkl<100> and \hkl<111> directions. Experimental values are $42\pm1$ eV for \hkl<100> and $44\pm1$ eV for \hkl<111>, obtained at a temperature $\leq 7$ K~\cite{maury_frenkel_1978}. In simulations, it is not obvious how to report the values for a given crystal direction due to the possibility of small angular deviations leading to large differences in the TDEs, either due the randomness of thermal and zero-point displacements or simply due to the anisotropy of the TDE surface~\cite{nordlund_molecular_2006,byggmastar_effects_2018}. In experiments, the electron beam is spreading in the sample, so the measurement always actually probes some angular interval around the principal direction. Without knowledge of the precise details of the experimental setup, it is very difficult to know what the magnitude of this spread is. In simulations, one has to choose a tolerance around the exact desired crystal direction (at 0 K with the GAP, the TDE at e.g. exactly \hkl<100> is significantly higher than a few degrees away from \hkl<100>). Using a 10$\degree$ tolerance, the minimum TDE values obtained with the GAP are $45.5\pm0.5$ eV for the \hkl<100> direction and $51.5\pm0.5$ eV for \hkl<111>, slightly higher than the experimental values. However, allowing for a 15$\degree$ tolerance, the minimum around the \hkl<111> direction becomes $47.5\pm0.5$ eV, but remains the same for the \hkl<100> direction. For the \hkl<110> direction, GAP predicts a TDE of $78\pm2$ eV. The \hkl<110> direction was not accessible from the experimental measurements, but good fits to the measured data were obtained by assuming values in the 70--80 eV range~\cite{maury_frenkel_1978}.

\subsection{Self-interstitial atoms and clusters}

Defects in the form of vacancies and self-interstitial atoms and their clusters have been extensively studied by density functional theory calculations in the literature. In particular, the recent papers by Ma and Dudarev~\cite{ma_universality_2019,ma_symmetry-broken_2019,ma_effect_2019} provides a comprehensive database of the energetics of single vacancies and SIAs in bcc metals, while Alexander et al.~\cite{alexander_ab_2016} in detail studied the energetics of SIA clusters. Most conveniently, they also used \textsc{vasp} with very similar input as we used when constructing the training database (the only noteworthy difference being a 12-electron PAW potential compared to 14 valence electrons in our DFT). Hence, we can rely on their results to be consistent with our training data and therefore use them to benchmark our GAP against defect properties.

\begin{figure}
 \centering
 \includegraphics[width=\linewidth]{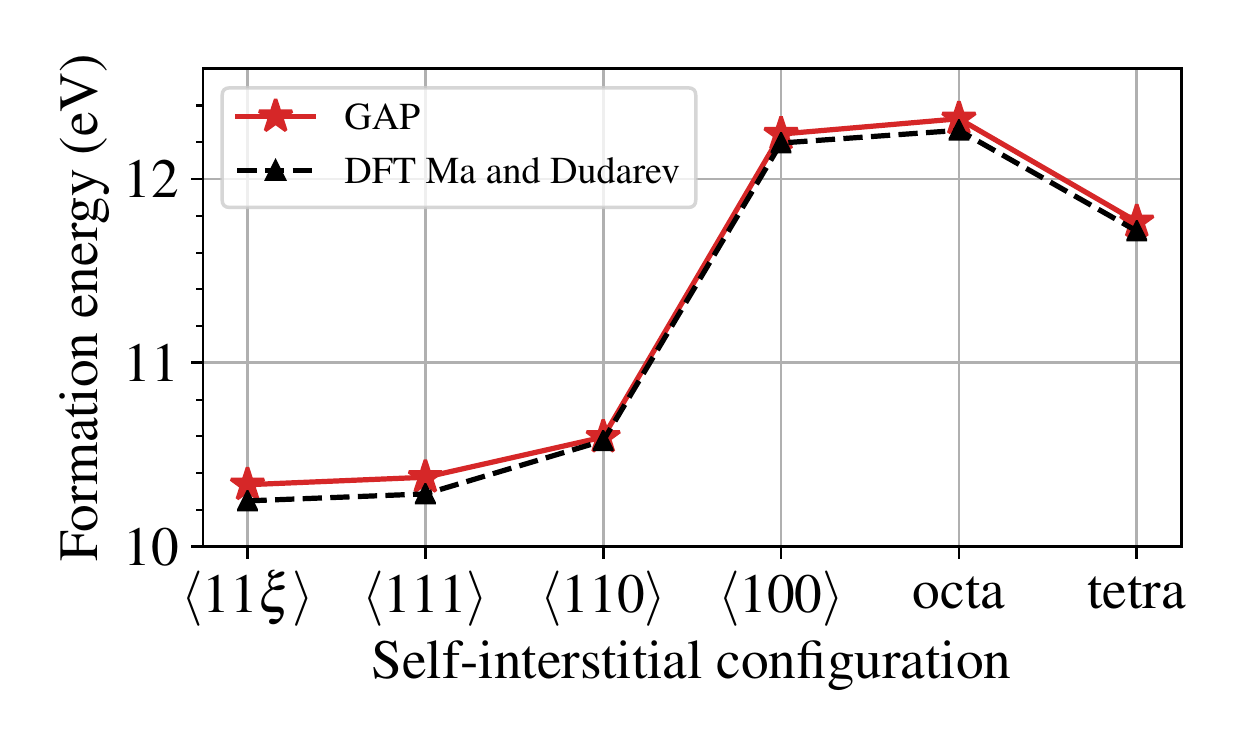}
 \caption{Formation energies of self-interstitial atoms in bcc W. The formation energy of the ground-state $\hkl<11\xi>$ is 10.34 eV in the GAP compared to 10.25 eV in DFT~\cite{ma_universality_2019}.}
 \label{fig:sia}
\end{figure}

\begin{figure}
 \centering
 \includegraphics[width=\linewidth]{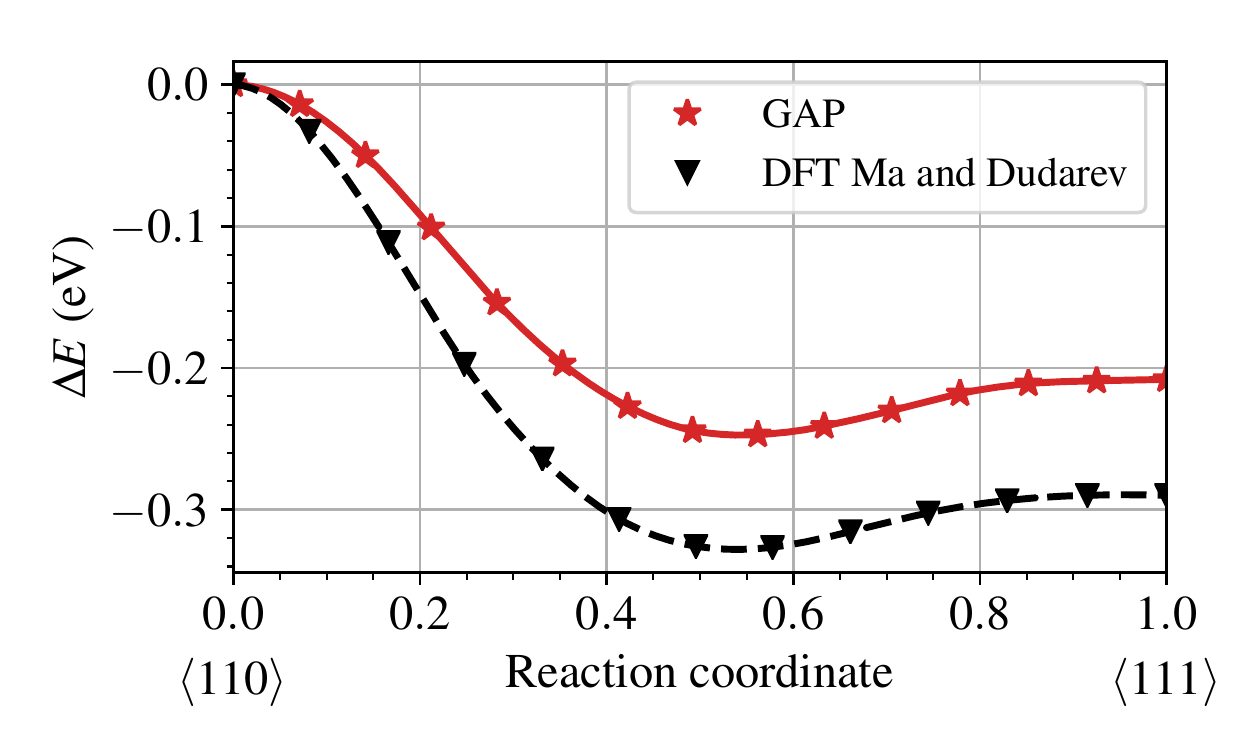}
 \includegraphics[width=\linewidth]{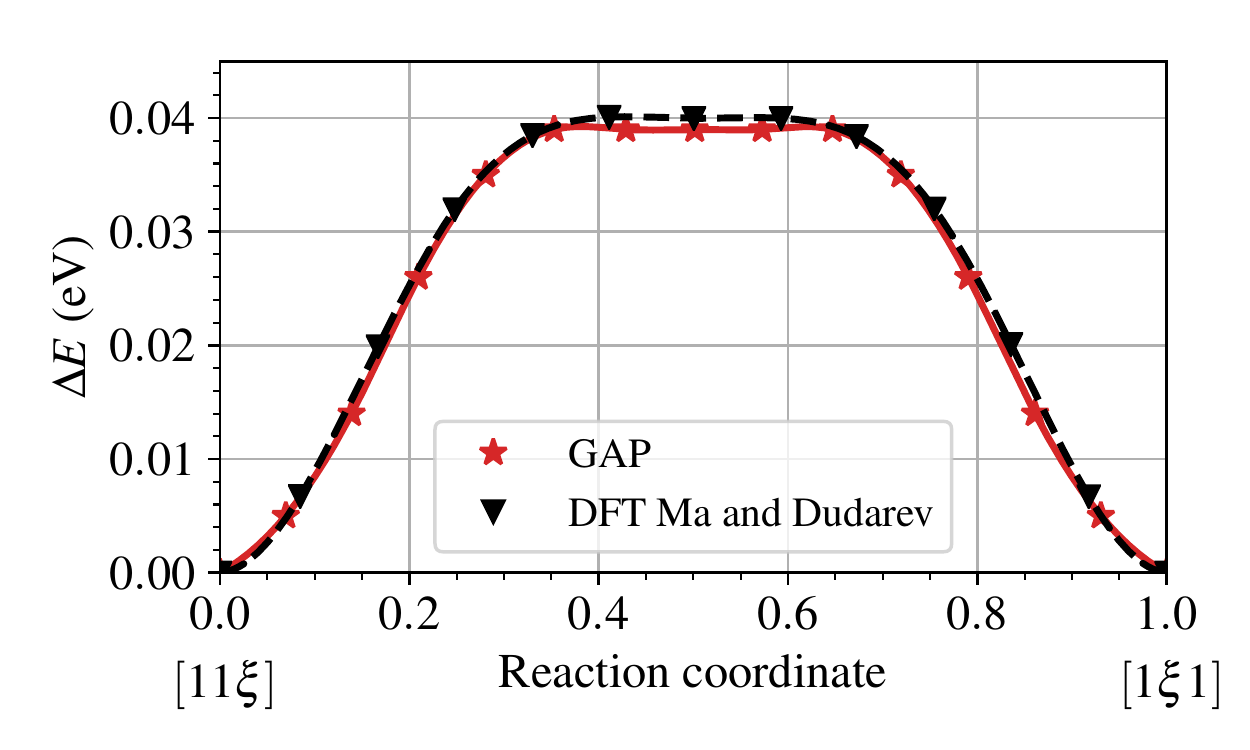}
 \caption{Top: Energy difference for a \hkl<110> dumbbell SIA rotating to the \hkl<111> direction, passing through the global minimum $\hkl<11\xi>$. Bottom: migration path between adjacent $\hkl<11\xi>$ configurations along the $\hkl<111>$ direction. DFT data are from~\cite{ma_symmetry-broken_2019}.}
 \label{fig:siamig}
\end{figure}

Single SIAs have historically been thought to stabilise as straight \hkl<111> dumbbells or crowdions, and migrate one-dimensionally through sequences of subtle \hkl<111> dumbbell-to-crowdion motion. However, it has been speculated~\cite{ventelon_ab_2012} and recently thoroughly demonstrated~\cite{ma_symmetry-broken_2019}, using DFT, that the most stable single SIA configuration in tungsten in fact is a tilted $\hkl<11\xi>$ configuration, where $\xi$ is close to 0.5. The difference in energy between the $\hkl<11\xi>$ and the straight \hkl<111> configuration is only 0.04 eV~\cite{ma_symmetry-broken_2019}. We did not explicitly include the $\hkl<11\xi>$ configuration in the training structures. Nevertheless, as we did sample various rotating dumbbells when constructing the training database, the GAP successfully reproduces the $\hkl<11\xi>$ configuration as the most stable single SIA. The difference in energy to the straight \hkl<111> dumbbell is 0.04 eV, consistent with DFT. Fig.~\ref{fig:sia} shows the formation energies of the common high-symmetry SIAs in bcc tungsten. The formation energies were calculated after minimising the positions and stress of a non-cubic box of 421 atoms, for which the elastic interactions across the periodic borders are minimal. The GAP formation energies are systematically around 0.1 eV higher than the DFT values from~\cite{ma_universality_2019}, except for the \hkl<110> dumbbell. Consequently, the difference in energy between the \hkl<111> and \hkl<110> configurations is only 0.21 eV, compared to 0.29 eV by DFT. This is also visible in Fig.~\ref{fig:siamig}, and might have consequences in high-temperature simulations, as the frequency \hkl<111>-to-\hkl<110> rotations will be overestimated. Despite efforts, we were not successful in eliminating this anomaly, which might be a consequence of the relatively small systems (121 atoms) included in the training database.

Fig.~\ref{fig:siamig} shows the main migration barriers of single SIAs calculated with the NEB method. The minimum along the \hkl<110>-to\hkl<111> rotation corresponds to the $\hkl<11\xi>$ configuration, with $\xi$ just above 0.5 in both GAP and DFT. Fig.~\ref{fig:siamig}b shows the expected zigzag migration path of a $\hkl[11\xi]$ SIA towards an adjacent $\hkl[1\xi1]$ position~\cite{ma_symmetry-broken_2019}. The GAP reproduces this migration barrier in excellent agreement with DFT. In addition to the static NEB calculations, we used the GAP to observe the migration of single SIAs in molecular dynamics simulations at low temperatures. We confirmed that it adopts the $\hkl<11\xi>$ symmetry and migrates in a one-dimensional zigzag-like manner along the path shown in Fig.~\ref{fig:siamig}b, consistent with the DFT-based predictions discussed in~\cite{ma_symmetry-broken_2019}.

Most existing interatomic potentials for radiation damage are fitted so that single SIAs are described well. However, it should also be transferable to larger clusters that readily form in e.g. collision cascade simulations. We found that fitting only single SIAs does not guarantee transferability to larger clusters, and that di-SIAs (both parallel and non-parallel dumbbell configurations) must be included in the training database. For tungsten, the majority of existing potentials struggle to reproduce the correct trend of the relative stability of clusters of multiple SIAs. For example, several widely used EAM potentials predict dislocation loops with the Burgers vector \hkl<100> to be lower in energy than the 1/2\hkl<111> loops~\cite{byggmastar_collision_2019}, which is in clear contradiction to DFT~\cite{alexander_ab_2016} and experimental observations~\cite{yi_situ_2013,yi_-situ_2016}. We therefore put particular focus on ensuring that our GAP reproduces the expected trend obtained by DFT.

\begin{figure}
 \centering
 \includegraphics[width=\linewidth]{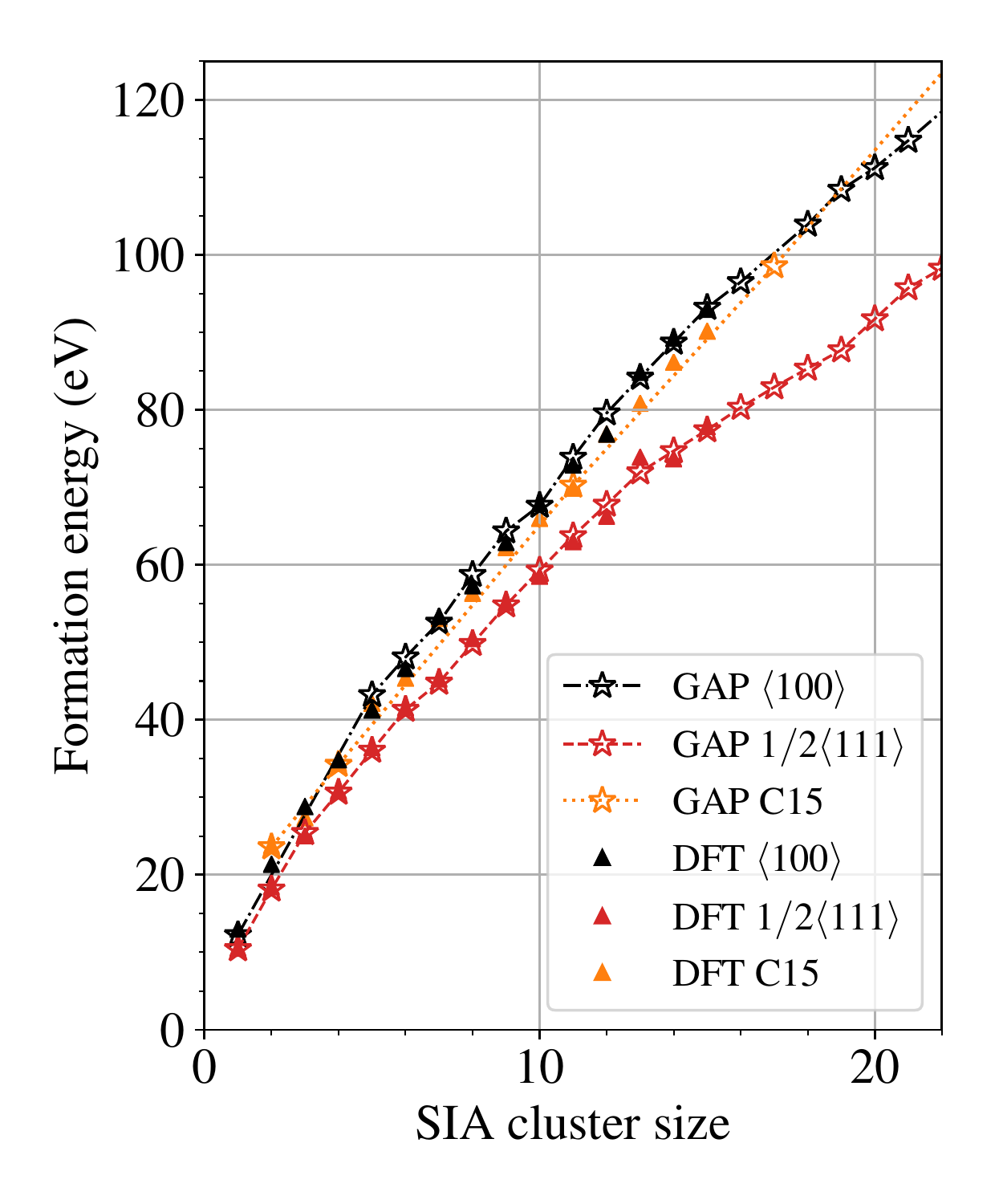}
 \caption{Formation energies of self-interstitial clusters in W compared between GAP and DFT data from Ref.~\cite{alexander_ab_2016}. Note that only sizes 1 and 2 were fit, so all the other data points serve as tests of the potential.}
 \label{fig:siaclust}
\end{figure}

Fig.~\ref{fig:siaclust} shows formation energies of parallel \hkl<111> and \hkl<100> SIA clusters (i.e. dislocation loops) compared between the GAP and DFT data from~\cite{alexander_ab_2016}. 1/2\hkl<111> clusters are created by inserting parallel dumbbells with a \hkl(110) habit plane, and \hkl<100> with a \hkl(100) plane, as in~\cite{alexander_ab_2016}. We also include the C15 clusters, which for small sizes have energies between the two dislocation loops. Overall, the GAP data closely overlaps with the DFT data across the entire DFT size range.

\subsection{Vacancies and vacancy clusters}

\begin{figure}
 \centering
 \includegraphics[width=\linewidth]{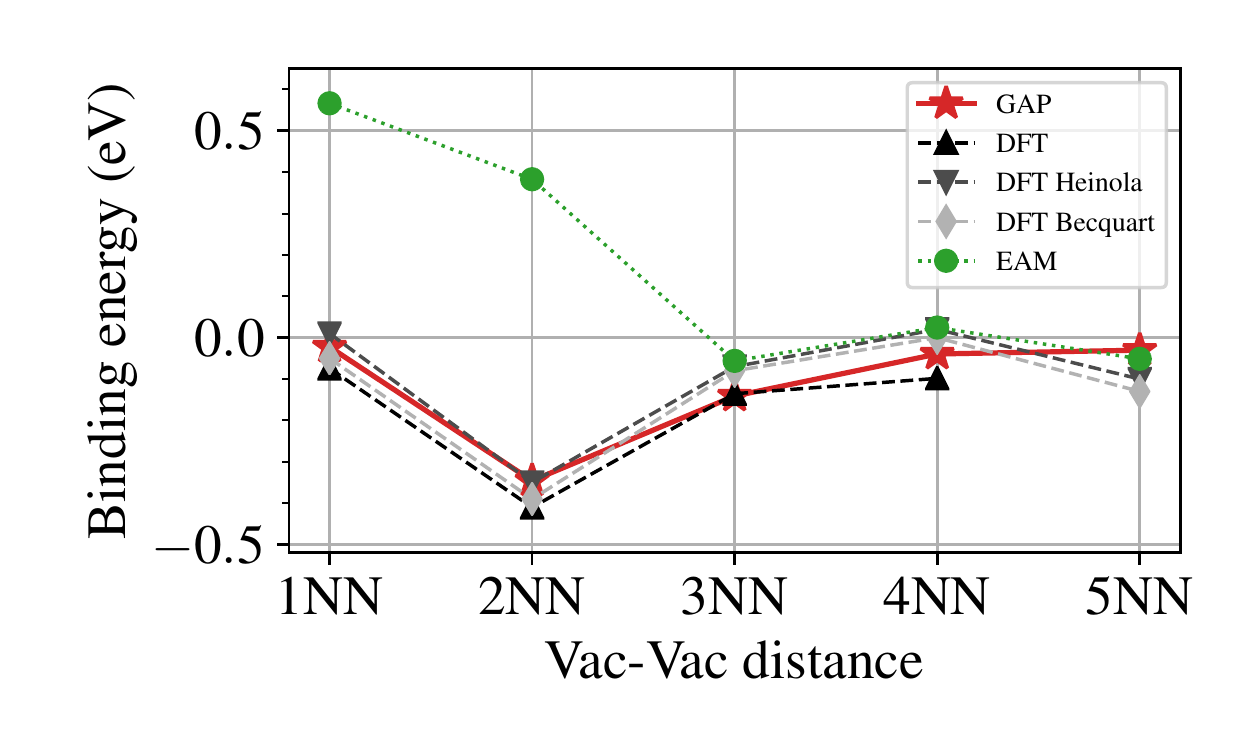}
 \caption{Binding energies of a di-vacancy in bcc W at different nearest-neighbour separations compared with DFT from Refs.~\cite{heinola_stability_2018,becquart_ab_2007} and our own. Only the 1NN and 2NN configurations were included in the training database. The EAM results are obtained using the potential from Ref.~\cite{derlet_multiscale_2007}, but is representative of the trend reproduced by most other EAM potentials as well.}
 \label{fig:divac}
\end{figure}

The vacancy formation energy and the vacancy migration barrier given by the GAP are consistent with DFT, as seen in Tab.~\ref{tab:bulk}. This is expected as both of these properties are well-represented by the training structures. The binding of di-vacancies is a peculiar feature of tungsten and some other bcc transition metals. DFT predicts that the binding energy of the second-nearest neighbour (2NN) di-vacancy is strongly repulsive, while other NN separations provide either weakly binding or weakly repulsive configurations, as shown in Fig.~\ref{fig:divac}. Reproducing this behaviour has presented a challenge for the vast majority of traditional interatomic potentials, but can be captured by the GAP as seen in Fig.~\ref{fig:divac}. Note that only the 1NN and 2NN di-vacancy configurations were included in the training database. Overall, the GAP reproduces the di-vacancy binding trend in good agreement with DFT, with only the 5NN configuration being slightly more stable than DFT predictions.

\begin{figure}
 \centering
 \includegraphics[width=\linewidth]{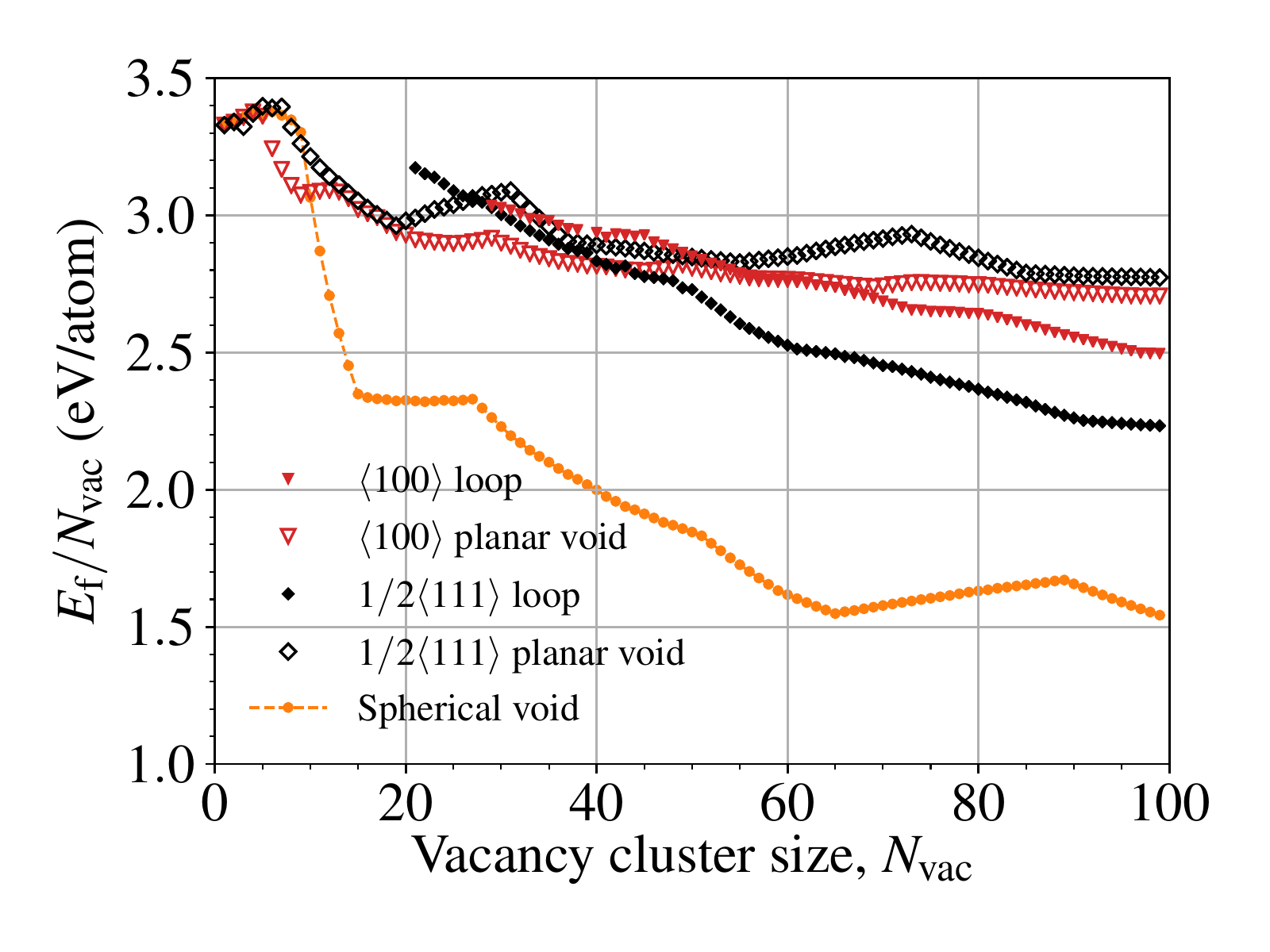}
 \caption{Formation energies of vacancy clusters in W obtained with the GAP.}
 \label{fig:vacclust}
\end{figure}

Larger clusters of vacancies form three-dimensional voids or planar dislocation loops. These include spherical voids, $1/2\hkl<111>$ and \hkl<100> dislocation loops. Calculations with existing potentials have shown that small vacancy dislocation loops are unstable and ''open up'' in the direction normal to the loop plane during relaxation~\cite{gilbert_structure_2008,fikar_nano-sized_2018}, forming what we refer to as planar voids. The critical sizes at which dislocation loops become more stable than their corresponding planar voids are, however, vastly different in different interatomic potentials. Some potentials predict the crossovers to occur already at a few tens of vacancies (1--2 nm diameters), while other predict crossovers at sizes well above 100 vacancies~\cite{fikar_nano-sized_2018}. Since the size of the simulation cell needs to be relatively large for clusters of this size, obtaining reliable DFT results to resolve this discrepancy is difficult, although efforts are currently ongoing~\cite{ma_unpublished_2019}. We used the GAP to investigate the relative stability of the different types of vacancy clusters. As the GAP is trained to di- and tri-vacancies and accurately reproduces surface energies, we expect it to be reasonably transferable to larger vacancy clusters.

We created \hkl<100> and $1/2\hkl<111>$ vacancy clusters by removing atoms in two or three consecutive \hkl<100> or \hkl<111> planes, respectively. To create dislocation loops, the surrounding atomic layers were compressed to create an initial strain field. For a cluster of size $N$ vacancies, the $N$ nearest atoms were removed in the corresponding planes for planar clusters, and in 3D for voids, resulting in clusters as close to circular and spherical shapes as possible. The simulation cells contained around 5500 atoms for clusters below 40 vacancies, and 16000 atoms for clusters in the 40--100 size range. Fig.~\ref{fig:vacclust} shows the formation energies per vacancy for the different clusters, calculated after a minimisation of the atomic positions and pressure.

The GAP predicts spherical voids to be the most stable vacancy cluster. The sharp local minima and maxima of the voids in Fig.~\ref{fig:vacclust} correspond to symmetric configurations. The $1/2\hkl<111>$ loop is the most stable planar configuration for sizes above around 40 vacancies, consistent with experimental observations of (both interstitial and vacancy) dislocations loops~\cite{yi_-situ_2016}. For dislocation loops, only energies of stable sizes are shown in Fig.~\ref{fig:vacclust}. Loops smaller than 20 vacancies for $1/2\hkl<111>$ and smaller than 30 vacancies for \hkl<100> spontaneously open up into planar voids during relaxation. The almost constant or slightly increasing formation energy per vacancy at small clusters seen in Fig.~\ref{fig:vacclust} is indicative of the weak or sometimes repulsive binding energies of small vacancy clusters in tungsten. The crossovers in stability between dislocation loops and planar voids occurs at 25 vacancies for $1/2\hkl<111>$ (but they remain very close in energy up to 40 vacancies) and at 55--60 vacancies for \hkl<100> clusters. This is roughly consistent with recent DFT results, which predicted a crossover at around 45 vacancies for $1/2\hkl<111>$ clusters~\cite{ma_unpublished_2019}.

\section{Conclusions and outlook}
\label{sec:concl}

We have shown that a machine-learning potential (GAP) with a moderately sized training database can capture a variety of properties of tungsten with essentially DFT accuracy. Even though the potential is fairly general, we particularly focused on reproducing properties relevant for radiation damage. The flexibility of the machine-learning framework allows the potential to describe properties that have been persistent challenges for analytical potentials, such as the relative stability of defect clusters and various surface properties. Hence, the potential will be useful for extracting more accurate data from classical molecular dynamics simulations of radiation damage in fusion-relevant tungsten, and settle previously unclear discrepancies in results with different existing potentials~\cite{byggmastar_collision_2019,fellman_radiation_2019}. We should, however, emphasise that the computational cost of the GAP with the current implementation is about 2--3 orders of magnitude higher than traditional analytical potentials. The high computational cost makes it challenging to obtain extensive statistics of radiation damage, but recent work on optimisation of the SOAP kernel has shown promising speed-ups without loss of accuracy~\cite{caro_optimizing_2019}.

The GAP also provides a good basis for further extension or development of potentials tailored to specific applications that are not reflected by our training structures. Additionally, the potential can be useful as a basis for extension to multi-component potentials, such as tungsten-based alloys or potentials for plasma-wall interactions in fusion reactor conditions. In the latter case, the accurate description of various surface reconstructions and surface energies provides an attractive basis for more accurate modelling of fusion-relevant W--H and W--He surface interactions (by adding analytical or machine-learned potentials for the light elements). Additionally, the training structures and fitting strategy can be easily repeated to develop similar potentials for other bcc metals. Efforts in these directions are ongoing and will be published elsewhere.

The potential files and the training database are available as supplementary material and from Ref.~\cite{gitlab_gap}.

\section*{Acknowledgements}

This work has been carried out within the framework of the EUROfusion Consortium and has received funding from the Euratom research and training programme 2014-2018 and 2019-2020 under grant agreement No 633053. The views and opinions expressed herein do not necessarily reflect those of the European Commission. Grants of computer capacity from CSC - IT Center for Science, Finland, as well as from the Finnish Grid and Cloud Infrastructure (persistent identifier urn:nbn:fi:research-infras-2016072533) are gratefully acknowledged. J.B. acknowledges helpful discussions with G. Csányi.

\clearpage
\appendix
\section{GAP predictions at extremely short interatomic distances}
\label{app:reppot}

\begin{figure}
 \centering
 \includegraphics[width=\linewidth]{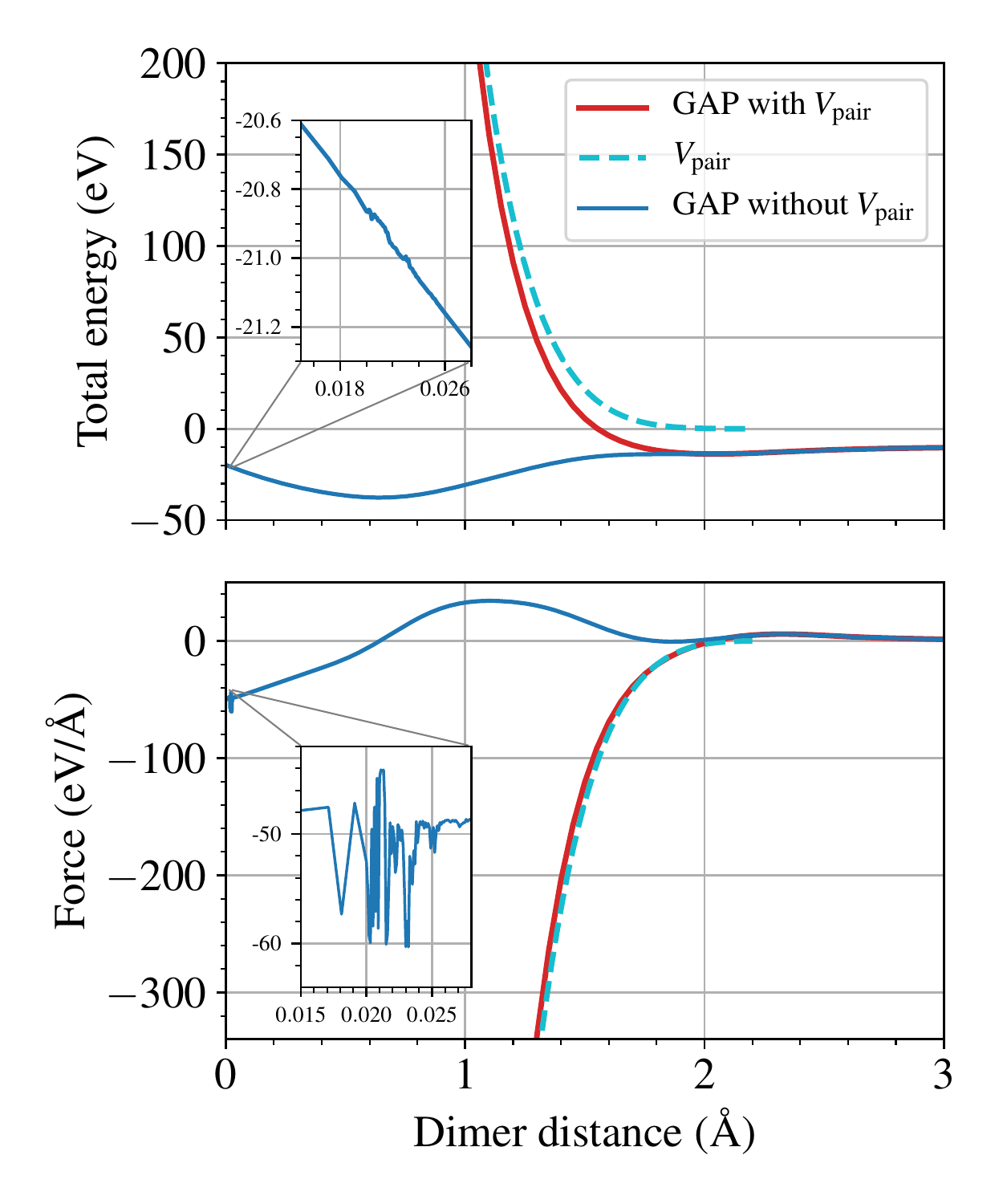}
 \caption{The energies and forces given by the GAP at short interatomic distances with and without the external pair potential. The zoomed-in insets show the numerical instability of GAP at extremely short interatomic distances.}
 \label{fig:puregap}
\end{figure}

The short-range part of the potential is dominated by the external screened Coulomb potential, as discussed in the main text. Nevertheless, it is crucial to make sure that the machine-learning extrapolations of the energies and forces at short interatomic distances do not interfere with the pair potential (i.e. remain smooth and negligible in magnitude). Fig.~\ref{fig:puregap} shows the energies and forces predicted by the GAP with and without the added pair potential for the dimer curve. Following the strategy described in the main text, the energies and forces given by GAP without the pair potential are negligible in comparison to the contributions from the pair potential, as desired. However, we found that the GAP becomes unstable at some distance close to zero, due to numerical limitations of the spherical harmonics expansion used in the SOAP descriptor. This is visible as kinks in the energy curve, leading to diverging forces as illustrated in the zoomed-in insets in Fig.~\ref{fig:puregap}. For previous GAPs for W, Fe, and Si~\cite{szlachta_accuracy_2014,dragoni_achieving_2018,bartok_machine_2018}, this instability occurs at distances in the 0.15--0.4 Å range, which might very well be reached in e.g. collision cascade simulations. Although none of the previous GAPs included a realistic repulsive part and are not suitable for cascade simulations, they can be made so by adding a repulsive pair potential.

A simple approach to eliminate the instability is to employ a smooth switching scheme between the GAP and a repulsive pair potential, similar to what is typically done with EAM and Tersoff-like potentials~\cite{bjorkas_comparative_2007} (although it becomes slightly less straight-forward due to the pure many-body nature of the SOAP descriptor). We tested such a scheme, in which the contributions of the GAP term is smoothly forced to zero, while the full screened Coulomb potential, $V_\mathrm{pair}$, remains present. The total energy of atom $i$ is then evaluated as
\begin{align}
 \begin{split}
    E_i &= S(i)\left[\sum_j V_\mathrm{pair} + E_\mathrm{GAP}(i)\right] + [1 - S(i)]\sum_j V_\mathrm{pair} \\
    &= S(i)E_\mathrm{GAP}(i) + \sum_j V_\mathrm{pair}, 
 \end{split}
\end{align}
where $j$ loops over all atoms within the cutoff range of atom $i$ and $S(i)$ is a switching function that depends on the environment of atom $i$ and goes smoothly to zero when the environment contains very short distances. In our test, we simply let $S(i) = S(r_\mathrm{min})$, where $r_\mathrm{min}$ is the shortest interatomic distance from atom $i$. For the switching function we chose the cutoff function in Eq.~\ref{eq:cutoff} (but inverted to approach zero as $r$ decreases). An almost identical approach was recently proposed for making deep learning neural network potentials applicable to irradiation simulations~\cite{wang_deep_2019}.

We found that our GAP becomes numerically unstable only below around 0.03 Å, as seen in Fig.~\ref{fig:puregap}. These distances will never be reached even in high-energy cascade simulations, since the pair potential contributes with energies in the MeV range. Hence the numerical instability is of no practical concern, and there is no need to employ the above switching scheme for our GAP. Nevertheless, we emphasise that when developing a GAP for radiation damage, it is crucial to ensure that the numerical limit of the SOAP implementation is beyond reach for any practical MD simulation, or eliminated by a switching scheme.

\bibliography{mybib}

\end{document}